\documentclass[prd,twocolumn,amsmath,amssymb,nofootinbib,floatfix,superscriptaddress]{revtex4}

\usepackage{graphicx,bm}
\usepackage[normalem]{ulem} 

\makeatletter
\def\graphicscale{\twocolumn@sw{0.3}{0.4}}
\def\graphicthreescale{\twocolumn@sw{0.3}{0.4}}

\begin{document}

\title{Three-dimensional lattice multiflavor scalar chromodynamics: \\
interplay between global and gauge symmetries}

\author{Claudio Bonati} 
\affiliation{Dipartimento di Fisica dell'Universit\`a di Pisa 
       and INFN Largo Pontecorvo 3, I-56127 Pisa, Italy}

\author{Andrea Pelissetto}
\affiliation{Dipartimento di Fisica dell'Universit\`a di Roma Sapienza
        and INFN Sezione di Roma I, I-00185 Roma, Italy}

\author{Ettore Vicari} 
\affiliation{Dipartimento di Fisica dell'Universit\`a di Pisa
       and INFN Largo Pontecorvo 3, I-56127 Pisa, Italy}

\date{\today}

\begin{abstract}
We study the nature of the finite-temperature transition of the
three-dimensional scalar chromodynamics with $N_f$ flavors. These
models are constructed by considering maximally O($M$)-symmetric
multicomponent scalar models, whose symmetry is partially gauged to
obtain SU($N_c$) gauge theories, with a residual nonabelian global
symmetry given by U($N_f$) for $N_c \ge 3$ and Sp($N_f)$ for $N_c=2$,
so that $M=2N_cN_f$.  We find that their finite-temperature transition
is continuous for $N_f = 2$ and for all values of $N_c$ we
investigated, $N_c=2,3,4$.  Such continuous transitions belong to
universality classes related to the global symmetry group of the
theory. For $N_c = 2$ it belongs to the SO(5)$ = $
Sp(2)/$\mathbb{Z}_2$ universality class, while for $N_c \ge 3$ it
belongs to the SO(3)$ = $ SU(2)/$\mathbb{Z}_2$ universality class.
For $N_f \ge 3$, the transition is always of first order.  These
results match the predictions obtained by using the effective
Landau-Ginzburg-Wilson approach in terms of a gauge-invariant order
parameter. Our results indicate that the nonabelian gauge degrees of
freedom are irrelevant at the transition. These conclusions are
supported by an analysis of gauge-field dependent correlation
functions, that are always short-ranged, even at the transition.
\end{abstract}

\maketitle


\section{Introduction}

The importance of symmetries in modern physics can be hardly
overestimated.  Global symmetries and the way in which they are
realized are commonly used to identify and describe different phases
of matter \cite{Anderson-book}. Local gauge symmetries play a
fundamental role both in particle physics, where they lie at the heart
of the Standard Model \cite{Weinberg-book}, and in condensed-matter
physics, where their applications span from superconductivity
\cite{Anderson-63} to topological order and quantum phase transitions
\cite{Sachdev-19}.

Several systems of physical interest display both global and local
symmetries, and a fundamental problem is to understand which of these
symmetries play a role in determining the universal behavior of the
system close to a continuous phase transition.  The traditional
Landau-Ginzburg-Wilson (LGW) approach to critical phenomena relies on
statistical field theory \cite{Landau-book, WK-74, Fisher-75, PV-02,
  ZJ-book}. In this scheme critical properties depend only on the
global symmetry breaking pattern and on some ``kinematic'' parameters,
like the space dimensionality and the number of fields components. For
transitions to/from topologically ordered states this time-honored
scheme has to be modified, due to the peculiar nonlocal character of
topological order \cite{SBSVF-04, WNMXS-17}.  However, when a
continuous phase transition emerges due to the breaking of a global
symmetry in a gauge theory, it is by no means obvious which is the
role played by the gauge degrees of freedom: do they affect the
critical behavior or not?

The study of the chiral phase transition in massless Quantum
Chromodynamics (QCD) was likely the first occasion in which this
problem could have been raised. Massless QCD is indeed invariant under
local SU(3) color transformations and under global
$\mathrm{SU}_L(N_f)\times \mathrm{SU}_R(N_f)$ flavor transformations,
with the chiral transition being associated with the symmetry breaking 
pattern
$\mathrm{SU}_L(N_f)\times \mathrm{SU}_R(N_f)\to
\mathrm{SU}_V(N_f)$~\cite{Weinberg-book}.
 However, starting from the seminal work of
Pisarski and Wilczek \cite{PW-84} (see Refs.~\cite{BPV-03, PV-13} for
some refinements), it was always implicitly assumed that gauge degrees
of freedom are irrelevant at the chiral transition, whose properties
were predicted by using a gauge-invariant order parameter and the LGW
approach.  Numerical lattice results later supported these
predictions, although with limited numerical precision because of the
computational burden of simulating dynamical fermions.  Moreover, in
recent times possible hints of discrepancies have appeared (see
Refs.~\cite{DElia:2018fjp, Sharma-19} for recent reviews).

The dependence of the critical behavior on the gauge degrees of
freedom can be numerically investigated much more accurately in scalar
models.  The three dimensional (3D) abelian case attracted much
attention in the recent past, both from the theoretical and from the
numerical point of view \cite{SBSVF-04, Nahum:2011zd, Nahum:2013qha,
  WNMXS-17, Pelissetto:2017sfd, Pelissetto:2017pxb, YHXV-18, YHVX-18,
  TS-18, PV-19, PV-19-2}. In particular, some works
\cite{Nahum:2011zd, Nahum:2013qha, Pelissetto:2017sfd,
  Pelissetto:2017pxb} reported some numerical evidence that the LGW
approach, based on a gauge invariant order parameter, may not describe
the emerging critical behavior.

Notwithstanding their applications to high-energy physics (most
notably to the QCD chiral phase transition but also to possible
extensions of the standard model) and their growing importance in
condensed-matter physics \cite{Sachdev-19, WNMXS-17, GASVW-18,
  SSST-19}, 3D nonabelian gauge theories has been so far much less
studied. The only case that was systematically investigated was that
of the 3D SU(2) gauge theory coupled to a scalar SU(2) doublet, which
is relevant for the electroweak phase transition (see, e.g.,
Refs.~\cite{Nadkarni:1989na, Kajantie:1993ag, Buchmuller:1994qy,%
  Kajantie:1996mn, Hart:1996ac}). For our purposes, however, this
model is somehow trivial, since it is known that its phase diagram
consists of a single phase \cite{OS-78, FS-79, DRS-80}.

To improve on this state of affairs, in Ref.~\cite{Bonati:2019zrt} we presented
results regarding a multiflavor 3D lattice scalar model with continuous
U($N_f$) symmetry, which might be called lattice multiflavor scalar
chromodynamics. We determined the transitions in this model, investigated their
nature, and compared the results with the predictions of two field-theoretical
formalisms, the gauge-invariant LGW scheme and the continuum scalar
chromodynamics.  The outcome of this analysis was that the LGW approach
correctly predicts the finite-temperature critical behavior of 3D multiflavor
scalar chromodynamics in all cases we studied, i.e., for $N_c=2, 3, 4$ and
$N_f=2,3$.  The analysis of the lattice results reported in  Ref.~\cite{Bonati:2019zrt}
 was however necessarily sketchy, and in this paper we
report all the analyses that permitted us to unambiguously identify the order
of the transitions and the universality class in the case of continuous
transitions. A more detailed discussion of the symmetries of the model,
and in particular of the U(1) flavor symmetry, is also reported, together with
the full details of the LGW approach for $N_c=2$, in which case the global
symmetry of the model is Sp($N_f$).

The paper is organized as follows. In Sec.~\ref{sec:model} the lattice
multiflavor scalar chromodynamics model is introduced, with a
discussion of its global and local symmetries. In Sec.~\ref{sec.III}
we discuss the predictions of the effective LGW approach. In
Sec.~\ref{sec:obs} we describe the lattice observables adopted and we
briefly summarize the finite-size scaling (FSS) results we use in the
analysis of the data.  In Sec.~\ref{sec:numerics} we present the
results of the numerical simulations, and finally we draw our
conclusions in Sec.~\ref{sec:concl}. In App.~\ref{app:symp} we discuss
the symplectic order parameters and, for $N_f = 2$, the relation
between Sp(2) and O(5) observables.  App.~\ref{app:sympLGW} is devoted
to a discussion of the LGW approach for the two-color case in which
the global symmetry group is Sp($N_f$). Finally, in
App.~\ref{app:largebeta} we discuss some properties of the model for
$\beta\to\infty$.

\section{The lattice model}
\label{sec:model}

The three-dimensional lattice model we are going to study has
$N_c\times N_f$ complex matrix variables $Z^{af}_{\bm x}$ associated
with each site ${\bm x}$ of a cubic lattice. Our starting point is the
lattice model defined by the action
\begin{align}
&S_{\rm inv} =- J \sum_{{\bm x},\mu} {\rm Re} \,
{\rm Tr} \,Z_{\bm x}^\dagger Z_{{\bm x}+\hat\mu}  +
\sum_{\bm x} V( {\rm Tr} Z^\dagger_{\bm x}\,Z_{\bm x})\, , \label{hiom}\\
&V(X) =  r\, X  +  u\, X^2\, .
\label{potential}
\end{align} 
In Eq.~\eqref{hiom} the first sum is over the lattice links, the
second one is over the lattice sites, and
 $\hat{\mu} =\hat{1},\hat{2},\hat{3}$ are unit vectors along the three lattice
directions. In particular, we consider the unit-length limit of the
site variables, which is formally obtained by setting $r = - u$, and
taking the limit $u\to\infty$ in the potential~(\ref{potential}), so
that the variables $Z$ satisfies
\begin{equation}
{\rm  Tr}\,Z_{\bm x}^\dagger Z_{\bm x} = 1\,, 
\end{equation}
and the action simplifies to 
\begin{eqnarray} \label{ullimit}
S_{\rm inv} = - J \sum_{{\bm x},\mu} {\rm Re} \,
{\rm Tr} \,Z_{\bm x}^\dagger Z_{{\bm x}+\hat\mu} \,. 
\end{eqnarray} 
Models with actions (\ref{hiom}) and (\ref{ullimit}) are invariant
under O($M$) transformations with $M = 2 N_c N_f$. This is immediately
checked if we write the matrices $Z_{\bm x}$ in terms of $M$-component
real vectors ${\bm S}_{\bm x}$. In the new variables, we obtain the
standard O($M$) nonlinear $\sigma$-model
\begin{eqnarray}
S_M = - J \sum_{{\bm x},\mu} {\bm S}_{\bm x}\cdot {\bm S}_{{\bm
    x}+\hat\mu}\,, \qquad {\bm S}_{\bm x} \cdot {\bm S}_{\bm x}=1\,.
\label{nvectorm}
\end{eqnarray} 

We now proceed by gauging some of the degrees of freedom. We associate
an SU($N_c$) matrix $U_{{\bm x},\hat{\mu}}$ with each lattice link and
extend the action (\ref{ullimit}) to ensure SU($N_c$) gauge
invariance. We also add a kinetic term for the gauge variables in the
Wilson form \cite{Wilson-74}. We obtain the model with action
\begin{equation}
\begin{aligned} 
S_g  &=- \beta N_f \sum_{{\bm x},\mu} 
{\rm Re}\, {\rm Tr} \left[ Z_{\bm x}^\dagger \, U_{{\bm x},\hat{\mu}}
\, Z_{{\bm x}+\hat{\mu}}\right] \\
& -
\frac{\beta_g}{N_c} \sum_{{\bm x},\mu>\nu} {\rm Re} \, {\rm Tr}\,
\left[
U_{{\bm x},\hat{\mu}} \,U_{{\bm x}+\hat{\mu},\hat{\nu}} 
\,U_{{\bm x}+\hat{\nu},\hat{\mu}}^\dagger  
\,U_{{\bm x},\hat{\nu}}^\dagger\right]
\,,
\end{aligned}
\label{hgauge}
\end{equation}
and partition function
\begin{equation}
Z = \sum_{\{Z,U\}} e^{-S_g}\,. \label{partfun}
\end{equation}
Note that the gauge group is SU($N_c$) and not U($N_c$), so that, for
$N_c=1$, the model is not related to the 3D CP$^{N_f-1}$ model
\cite{ZJ-book} or to the abelian Higgs model, studied, e.g., in
Ref.~\cite{PV-19-2}.  The factor $N_f$ in the first term is introduced
so that the large-$N_f$ limit can be performed by keeping $\beta$
fixed; the factor $1/N_c$ in the second term is instead conventional
in the lattice gauge theory literature. Note that, for
$\beta_g\to\infty$, the product of the gauge fields along a plaquette
converges to one, and therefore we can set $U_{{\bm x},\hat\mu} = 1$ modulo a
gauge transformation. Therefore, in this limit we reobtain the O($M$)
invariant theory (\ref{ullimit}) we started from.

It is immediate to see that, for any value of $N_c$ and $N_f$,
 $S_g$ is
invariant under the local gauge transformation
\begin{equation}
\label{gaugeinv}
Z_{\bm x}\to G_{\bm x} Z_{\bm x}\ , \quad U_{\bm x,\hat{\mu}}\to
G_{\bm x} U_{\bm x,\hat{\mu}} G_{\bm{x}+\hat{\mu}}^{\dag}\ ,
\end{equation}
with $G_{\bm x}\in $ SU($N_c$), and under the global transformation
\begin{equation}\label{flavinv}
Z_{\bm x}\to Z_{\bm x} V\ ,
\quad U_{\bm x,\hat{\mu}}\to U_{\bm x,\hat{\mu}} \ ,
\end{equation}
with $V\in $ U($N_f$). Note that, more precisely, the global symmetry group 
of the model is U($N_f$)$/{\mathbb Z}_{N_c}$, where ${\mathbb Z}_{N_c}$ is 
the center of the gauge symmetry group SU($N_c$).

Actually, for $N_c=2$ the action $S_g$ is invariant under a larger
global symmetry group, the compact complex symplectic
group\footnote{Several notations are used to denote this group: in
  particular both Sp($N_f$) and Sp(2$N_f$) can be found in the
  literature.} Sp($N_f$).  This is a well established result (we found
mention of it, in various forms, e.g., in
Refs.~\cite{Georgi-book,AY-94,DP-14,WNMXS-17}), which is a consequence
of the self-duality of the fundamental representation of SU(2).  We
will here briefly explain the origin of this symmetry enlargement,
introducing also some notations that will be useful in the following.

We define 
\begin{equation}
Y_{\bm x}^{af} = \sum_b \epsilon^{ab} \bar{Z}_{\bm x}^{bf},
\end{equation}
where $\epsilon^{ab}$ is the completely antisymmetric tensor in 2 dimensions
($\epsilon^{12} = - \epsilon^{21} = 1$), 
and the $2\times 2N_f$ matrix field
$\Gamma_{\bm x}^{a \alpha}$, defined by
\begin{equation}
\Gamma_{\bm x}^{a \alpha}=
\left\{ \begin{array}{lll} 
Z_{\bm x}^{a \alpha} & \mathrm{if} & 1\le \alpha \le N_f \\
Y_{\bm x}^{a\, (\alpha-N_f) } & \mathrm{if} & N_f+1 \le \alpha\le 2 N_f 
\end{array}\right.\ . 
\label{gammadef}
\end{equation}
Since SU(2) matrices satisfy
\begin{equation}
\sum_b \epsilon^{ab} \bar{U}^{bc} = \sum_b U^{ab} \epsilon^{bc},
\end{equation}
$\Gamma_{\bm x}$ transforms covariantly
under gauge transformations: 
\begin{equation}
\Gamma_{\bm x} \to G_{\bm x} \Gamma_{\bm x} \, .
\end{equation}
We can now rewrite 
the nearest-neighbor interaction term involving the scalar variables as
\begin{equation}
\begin{aligned} 
& \frac{1}{2}\sum_{f,a,b}\left[ \bar{Z}_{\bm x}^{af} \, 
U_{{\bm x},\hat{\mu}}^{ab}\, Z_{{\bm x}+\hat{\mu}}^{bf}
+ Z_{\bm x}^{af} \, \bar{U}_{{\bm x},\hat{\mu}}^{ab}\, \bar{Z}_{{\bm x}
+\hat{\mu}}^{bf} \right] =  \\
& \frac{1}{2} \sum_{f,a,b}\left[ \bar{Z}_{\bm x}^{af} \, 
U_{{\bm x},\hat{\mu}}^{ab}\, Z_{{\bm x}+\hat{\mu}}^{bf}
+ \bar{Y}^{af}_{\bm x} \,\, U_{{\bm x},\hat{\mu}}^{ab}\, 
Y_{{\bm x}+\hat{\mu}}^{bf}\right]  =  \\
& \frac{1}{2}\sum_{\gamma,a,b} \bar{\Gamma}_{\bm x}^{a\gamma}
\, U_{{\bm x},\hat{\mu}}^{ab} \Gamma_{{\bm x}+\hat{\mu}}^{b\gamma} 
=\frac{1}{2}{\rm Tr}\,\Gamma^{\dag}_{\bm x} U_{{\bm x},\hat{\mu}}
\Gamma_{{\bm x}+\hat\mu}
\,.
\end{aligned}
\label{equico}
\end{equation}
Apparently, the action (\ref{equico}) 
is invariant under the global transformations
\begin{equation}
\label{gamtra}
\Gamma_{\bm x} \to \Gamma_{\bm x}M\ ,\quad M\in {\rm U}(2N_f)\, . 
\end{equation}
However, one should bear in mind that the $\Gamma$ variables are not
generic, since they are obtained by a formal doubling of the degrees
of freedom.  Therefore, one must only consider transformations $M$
that maintain the particular structure (\ref{gammadef}).  To identify
them, we note that the previous bipartite structure of $\Gamma$ is
equivalent to the relation
\begin{equation}
\sum_a \epsilon^{ab} 
\bar{\Gamma}_{\bm x}^{b\alpha} = - \sum_\gamma \Gamma_{\bm x}^{a\gamma} 
   J^{\gamma\alpha}\ ,
\label{condconj}
\end{equation}
where $J$ is the $2 N_f\times 2N_f$ matrix
\begin{equation}
J=\left(\begin{array}{cc} 0 & -I \\  I & 0\end{array}\right) \, , 
\end{equation}
and $I$ is the $N_f\times N_f$ identity matrix. Therefore, the global
invariance group of $S_g$ is the subgroup of U($2N_f$) which leaves
invariant the relation Eq.~(\ref{condconj}). By straightforward
manipulations it is possible to show that this requires $M$ to satisfy
\begin{equation}
M J M^T = J\, ,
\label{mscond}
\end{equation}
which identifies the global symmetry group as the compact (unitary)
complex symplectic group Sp($N_f$) (see, e.g.,
Ref.~\cite{Simon-book}).  The global symmetry group for $N_c=2$ is
thus Sp($N_f$)/${\mathbb Z}_2$, since the sign of the field can be
redefined by a gauge transformation.  Note that, for $N_f = 2$, we
have the isomorphism (see, e.g., Ref.~\cite{Simon-book})
\begin{equation}
\label{so5sp}
{\rm SO}(5)={\rm Sp}(2)/{\mathbb Z}_2\ .
\end{equation}
Finally, let us explicitly note that the Sp($N_f$) symmetry also holds
when the fields do not satisfy the unit-length condition.  Since
\begin{eqnarray}
\mathrm{Tr}\,Z_{\bm x}^\dagger Z_{\bm x}= 
\frac{1}{2} \mathrm{Tr}\,\Gamma_{\bm x}^\dagger \Gamma_{\bm x}\ ,
\label{zga}
\end{eqnarray}
is invariant under any U($2N_f$) transformations, and, in particular,
under those of its Sp($N_f$) subgroup, the action is Sp($N_f$)
invariant for  generic site potentials $V$ in Eqs.~(\ref{hiom})
and  (\ref{potential}).

\section{Effective field theory results} \label{sec.III}

The critical behavior of the lattice multiflavor scalar chromodynamics
was discussed in Ref.~\cite{Bonati:2019zrt}. Two different approaches
were considered: the continuum theory corresponding to the lattice
model and the Landau-Ginzburg-Wilson theory built in term of a
gauge-invariant order parameter. The renormalization-group flow of
continuum multiflavor chromodynamics was studied in the
$\varepsilon$-expansion around four dimensions~\cite{Das-18}. It was
found that a stable fixed point (FP) only exists for a very large
number of flavors [for $N_c=2$ it exists only for
  $N_f>359+O(\varepsilon)$]. As a consequence, for small values of
$N_f$ a first-order transition is predicted.

In the LGW approach, one starts by considering an order parameter that
breaks the global symmetry of the model. We first consider the case
$N_c > 2$, so that the global symmetry is U($N_f$)/${\mathbb
  Z}_{N_c}$. Since this is not a simple group, we may have different
symmetry breakings.

One possibility is that of breaking the SU$(N_f)$ subgroup.  An
appropriate order parameter is the field combination
\begin{equation}
Q^{fg}_{\bm x} = \sum_a {\bar Z}^{af}_{\bm x} Z^{ag}_{\bm x} - 
     {\delta^{fg}\over N_f}\,, 
\label{qdef}
\end{equation}
which is the natural generalization of the quantity studied in abelian
models (see, e.g., Refs.~\cite{PV-19, PV-19-2}).  The corresponding
LGW theory is obtained by considering a hermitian traceless $N_f\times
N_f$ matrix field $\Psi({\bm x})$, 
which represents a coarse-grained version of $Q_{\bm x}$,
with Lagrangian
\begin{eqnarray}
&&{\cal L}_{\rm LGW} = \hbox{Tr }\partial_\mu \Psi \partial_\mu \Psi + 
    r\, \hbox{Tr } \Psi^2  
\label{SLGW} \\ 
&& \qquad + \,u_3\, \hbox{Tr } \Psi^3 + u_{41} \,\hbox{Tr } \Psi^4 + 
      u_{42}\, (\hbox{Tr } \Psi^2)^2\,.
\nonumber
\end{eqnarray}
This Lagrangian is invariant under the global transformations
$\Psi \to V \Psi V^\dagger$ and
therefore the symmetry group is SU($N_f$)/${\mathbb Z}_{N_f}$. As
discussed in, e.g., Ref.~\cite{PV-19}, the cubic term vanishes for
$N_f = 2$. In this case a continuous transition is possible in the
SU(2)/${\mathbb Z}_2$, that is in the vector SO(3), universality
class.  For $N_f > 2$ the cubic term is present and, on the basis of
the usual mean-field arguments, one expects a first-order transition.

A second possibility is that of breaking the U(1)/$\mathbb{Z}_{N_c}$
symmetry group associated with the transformations
\begin{equation}
Z^{af}_{\bm x} \to e^{i\alpha} Z^{af}_{\bm x},
\end{equation}
which leave invariant the order parameter $Q_{\bm x}^{ab}$. 
However,  for $N_f < N_c$, this additional symmetry is only
apparent. Indeed, for any $\bm x$, one can find an SU($N_c$) matrix
$G_{\bm x}$ such that
\begin{equation}
     e^{i\alpha} Z_{\bm x} = G_{\bm x} Z_{\bm x}.
\label{relU1}
\end{equation}
If $N_f < N_c$, there is a gauge transformation ${Z'}_{\bm x} =
G_{1\bm x} Z_{\bm x}$ such that ${Z'}_{\bm x}^{af} = 0$ for any $f$
and any $a$ satisfying $N_f+1\le a \le N_c$. Then, one defines the
$N_c\times N_c$ unitary matrix
\begin{equation}
  G_2 = \hbox{diag }(g_1,\ldots,g_{N_c}) \qquad
\end{equation}
with $g_a = e^{i\alpha}$ for $1 \le a \le N_f$, 
$g_a = e^{-i\alpha N_f}$ for $a =N_f+1$, $g_a = 1$ for $a > N_f + 1$. 
It is then easy to verify that 
$G_x = G_{1\bm x}^\dagger G_2 G_{1\bm x}$ satisfies Eq.~(\ref{relU1}). 

For $N_f\ge N_c$, the relation (\ref{relU1}) does not hold anymore, and one must
consider the breaking of the abelian symmetry U(1)/${\mathbb
  Z}_{N_c}$. An appropriate order parameter is
\begin{equation}
D^{f_1,\ldots, f_{N_c}}_{\bm x} = 
   \sum_{a_1,\ldots,a_{N_c}} \epsilon^{a_1,\ldots,a_{N_c}} 
    Z^{a_1 f_1}_{\bm x}\ldots Z^{a_{N_c} f_{N_c}}_{\bm x},
\label{Ddef}
\end{equation}
which is invariant under gauge transformations (here
$\epsilon^{a_1,\ldots,a_{N_c}}$ is the completely antisymmetric tensor
in $N_c$ dimensions). Such an order parameter vanishes for $N_f <
N_c$, an expected result given the effective absence of the symmetry
in this case. For $N_f=N_c$ the order parameter defined in
Eq.~(\ref{Ddef}) is invariant under SU($N_f$) transformations and
therefore it is a good order parameter for the breaking of the U(1)
flavor symmetry.  It can be rewritten in a simpler way, as
\begin{equation}
D^{f_1,\ldots, f_{N_c}}_{\bm x} = 
\epsilon^{f_1,\ldots, f_{N_c}} \, \hbox{det } Z_{\bm x}.
\label{DdenNcNf}
\end{equation}
On the other hand, for $N_f > N_c$, the order parameter belongs to a
nontrivial representation of SU($N_f$). Therefore, it condenses only
if both the SU($N_f$) and the U(1) symmetries are broken.

As we discuss in App.~\ref{app:largebeta}, in our model, for $N_c\ge
3$, the order parameter $D^{f_1,\ldots, f_{N_c}}_{\bm x}$ vanishes for
$\beta\to\infty$. If we assume that the relevant configurations in the
low-temperature phase are simply obtained by considering short-range
fluctuations on top of the ordered background observed for $\beta =
+\infty$, we conclude that $D$-correlations are short-ranged in the
low-temperature phase, i.e., that the U(1) symmetry is not
broken. Below we will present numerical results for $N_c=N_f=3$ that
confirm this picture.

For $N_c = 2$ the symmetry group is Sp($N_f$)/${\mathbb Z}_2$. 
The order parameter is a symplectic analogue of 
$Q_{\bm x}$. Specifically, we define
\begin{equation}
{\cal T}_{\bm x}^{\alpha\beta} =
\sum_a \overline{\Gamma}_{\bm x}^{a\alpha} \Gamma_{\bm x}^{a\beta} -
     \frac{\delta^{\alpha\beta}}{2 N_f} \sum_{a\gamma}
     \overline{\Gamma}_{\bm x}^{a\gamma} \Gamma_{\bm x}^{a\gamma}\,,
\label{Tdef}
\end{equation}
with $\Gamma_{\bm x}^{a\alpha}$ defined in Eq.~\eqref{gammadef}. This
order parameter is a $2N_f\times 2N_f$ hermitian traceless matrix
which satisfies the additional condition
\begin{equation}
J\bar{\cal T}J+{\cal T}=0\ ,
\label{Spalgebra}
\end{equation} 
which follows from Eq.~\eqref{condconj}.  The matrix ${\cal T}$ is
thus an element of the $\mathfrak{sp}$($N_f$) algebra
\cite{Simon-book}.  The explicit construction of the corresponding LGW
theory starts by defining a $2N_f\times 2N_f$ hermitian traceless
matrix field $\Psi({\bm x})$ that satisfies the analog of
Eq.~(\ref{Spalgebra}). The corresponding LGW theory is obtained by
considering the most general quartic polynomial in the fields: we
reobtain Eq.~(\ref{SLGW}). For $N_f = 2$, as discussed in
App.~\ref{app:sympLGW}, the cubic term vanishes. Therefore, continuous
transitions are allowed in the SO(5) universality class, given the
isomorphism between Sp(2)/${\mathbb Z}_2$ and the SO(5) group.  For
$N_f > 2$, a cubic operator is generically present and therefore we
expect first-order transitions. Note, that for $N_c = 2$, there is no
residual U(1) symmetry, as U(1) global transformations are a subgroup
of the Sp($N_f$) group.

We finally note that the LGW approach based on the symmetry of the
model does not depend on the specific form of the lattice potential
$V(X)$ in Eq.~(\ref{potential}). Moreover, we recall that the presence of a stable
fixed point in the corresponding LGW theory does not exclude the
possibility that the model undergoes a first-order transition, when
the system is outside the attraction domain of the stable fixed point
even though it shares the global symmetry of the universality class.

\section{Observables and analysis method}
\label{sec:obs}

In this section we introduce the lattice observables studied and we
briefly recall some basic facts about FSS that will be relevant for
the analysis of the numerical data.  We always assume the lattice to
have periodic boundary conditions and to be of linear size $L$.

\subsection{Lattice observables} \label{sec4.1}

In the following we consider the energy density and the specific heat,
defined as
\begin{eqnarray}\label{ecvdef}
E = \frac{1}{\beta N_f V} \langle S_g \rangle\,,\quad
C =\frac{1}{\beta^2 N_f^2 V}\left( \langle S_g^2 \rangle 
- \langle S_g  \rangle^2\right)\,,\quad
\end{eqnarray}
where $V=L^3$. We also define the average gauge energy as 
\begin{equation}
E_g = {1\over 6 V N_c} \left \langle 
\sum_{{\bm x},\mu>\nu} {\rm Re} \, {\rm Tr}\,
\left[
U_{{\bm x},\hat{\mu}} \,U_{{\bm x}+\hat{\mu},\hat{\nu}}
\,U_{{\bm x}+\hat{\nu},\hat{\mu}}^\dagger
\,U_{{\bm x},\hat{\nu}}^\dagger\right] \right\rangle.
\label{gauge-energy}
\end{equation}
To study the breaking of the SU($N_f$) flavor symmetry we consider the
order parameter $Q$ defined in Eq.~(\ref{qdef}), which is a hermitian
and traceless $N_f\times N_f$ matrix. Its two-point correlation
function is defined by
\begin{equation} \label{gxyp}
G({\bm x}-{\bm y}) = \langle {\rm Tr}\, Q_{\bm x} Q_{\bm y} \rangle\,,  
\end{equation}
where the translation invariance of the system has been explicitly
taken into account. We can define the correasponding susceptibility
$\chi$ and correlation length $\xi$ as
\begin{align}
&\chi=\sum_{\bm x} G({\bm x}) \,,\label{chidef}\\ &\xi^2 = \frac{1}{4
    \sin^2 (\pi/L)} \frac{\widetilde{G}({\bm 0}) - \widetilde{G}({\bm
      p}_m)}{\widetilde{G}({\bm p}_m)}\,,
\label{xidefpb}
\end{align}
where $\widetilde{G}({\bm p})=\sum_{{\bm x}} e^{i{\bm p}\cdot {\bm x}} G({\bm
x})$ is the Fourier transform of $G({\bm x})$ and ${\bm p}_m = (2\pi/L,0,0)$.
We also consider the Binder
parameter $U$, defined by
\begin{equation}
U = \frac{\langle \mu_2^2\rangle}{\langle \mu_2 \rangle^2} \,, \qquad
\mu_2 = \frac{1}{V^2}  
\sum_{{\bm x},{\bm y}} {\rm Tr}\,Q_{\bm x} Q_{\bm y}\,.
\label{binderdef}
\end{equation}
We will study the U(1) flavor symmetry only for $N_f = N_c$. In this
case it is equivalent to consider the scalar order parameter, see
Eq.~(\ref{DdenNcNf}),
\begin{equation}
D_{\bm x} = \hbox{det }Z_{\bm x}\; .
\label{Ddefinition}
\end{equation} 
We define the correlation function
\begin{equation}
G_D({\bm x-y}) = \langle \hbox{Re}\,{\bar D}_{\bm x} D_{\bm y} \rangle\,,
\label{GDdef}
\end{equation}
the correlation length $\xi_D$ using the analogue of Eq.~(\ref{xidefpb}), and 
the Binder parameter 
\begin{equation}
U_D = \frac{\langle \mu_{D2}^2\rangle}{\langle \mu_{D2} \rangle^2}\, , \qquad
\mu_{D2} = \frac{1}{V^2}  
\sum_{{\bm x},{\bm y}} {\rm Re}\, \bar{D}_{\bm x} D_{\bm y}\,.
\label{binderDdef}
\end{equation}

To better appreciate the role of the gauge degrees of freedom, we also
study some observables involving the SU($N_c$) gauge link
variables. More specifically, we consider the averages
\begin{equation}
\left\langle  
  \sum_{ab} \bar{Z}_{\bm x}^{af}\, \left [\prod_{\ell \in \cal C} U_\ell \right]^{ab}\,
Z_{\bm y}^{bg} \right\rangle,
\end{equation}
where the product extends over the link variables that belong to a
lattice path $\cal C$ connecting the points $\bm x$ and $\bm y$. To
define quantities that have the correct FSS, the path ${\cal C}$ must
be chosen appropriately~\cite{APV-08}, and here we consider
correlations between points along lattice lines:
\begin{eqnarray} \label{Gv} 
G_V(t,L) = \hbox{Re}\left\langle \sum_{abfg} \bar{Z}_{\bm x}^{af}
\left[\prod_{k=0}^{t-1} U_{{\bm x}+k \hat{\mu},\hat\mu} \right]^{ab}
Z^{bf}_{{\bm x}+t\hat{\mu}} \right\rangle\,.\quad
\end{eqnarray}
As usual, translation invariance and independence of the direction
$\hat{\mu}$ can be used to actually increase the statistics. In some
test cases we also determined the Polyakov loop
\begin{equation}
   P(L) = \frac{1}{3 L^3} \sum_{{\bm x},\mu}
   \hbox{Re } \left\langle {\rm Tr}\,\left[
  \prod_{k=0}^{L-1}
 U_{{\bm x}+k\hat{\mu},\hat{\mu}}\right]
   \right\rangle\,.
   \label{Polyakov}
\end{equation}
For $N_f = 2$ and $N_c = 2$ the model is invariant under
Sp(2)/${\mathbb Z}_2 = O(5)$ transformations. We discuss in
App.~\ref{app:symp} the O(5) observables that can be defined in terms
of the order parameter (\ref{Tdef}). In particular, we show that the
second-moment correlation length computed from $G({\bm x})$, $G_D({\bm
  x})$ or the O(5)-invariant correlation frunction of the the order
parameter ${\cal T}^{\alpha\beta}$ are numerically the same. For the
Binder parameters, instead, the relation is not trivial. We have
\begin{equation}
U = {25\over 21} U_\Gamma\,, \qquad 
U_D = {10\over 7} U_\Gamma\,,
\label{relationsBinder22}
\end{equation}
where $U_\Gamma$ is the O(5)-invariant Binder parameter defined in
App.~\ref{app:symp}, which corresponds to the usual vector parameter
in the O(5) theory.

For $N_f=2$ and $N_c\ge 3$, the global symmetry group is
SU(2)/${\mathbb Z}_2 = SO(3)$. This invariance can be more easily
understood by defining the gauge-invariant three-component real vector
variables $\varphi^k_{\bm x}$ as
\begin{eqnarray} 
\label{vecvar}
\varphi_{\bm x}^k = 
\sum_{a,f, g} \bar{Z}_{\bm x}^{af} \sigma^k_{fg} Z_{\bm x}^{ag}
=\sum_{f, g}\sigma^k_{fg}Q_{\bm x}^{fg}\,, 
\end{eqnarray}
where $\sigma^k$ are the Pauli matrices. Previoulsy defined
observables, like $\chi$ and $U$, can be rewritten in term of the
vector variable ${\bm \varphi}_{\bm x}$ using
\begin{align}
&G({\bm x}-{\bm y}) = 
\frac{1}{2} \langle {\bm \varphi}_{\bm x}\cdot {\bm \varphi}_{\bm y}\rangle\,,
\label{gxphik}\\
&U = \frac{\langle \mu_2^2\rangle}{\langle \mu_2 \rangle^2}\, , \qquad
\mu_2 = \frac{1}{V^2} \sum_{{\bm x},{\bm y}} 
{\bm \varphi}_{\bm x}\cdot {\bm \varphi}_{\bm y}
\, .\label{uphik}
\end{align}
Note however that the vectors ${\bm \varphi}_{\bm x}$ do not have
fixed length, indeed
\begin{equation}
{\bm \varphi}_{\bm x}\cdot {\bm \varphi}_{\bm x}
  = 2\, {\rm Tr} \,Q_{\bm x}^2\le 1.
\label{phiknorm}
\end{equation}

\subsection{Finite-Size Scaling}

To investigate continuous transitions using FSS it is particularly
convenient to study RG invariant quantities, such as $U$ and
\begin{equation}\label{rxidef}
R_{\xi}=\xi/L\ .
\end{equation}
For an RG-invariant quantity, generically denoted by $R$, FSS theory predicts
the scaling behavior \cite{PV-02}
\begin{align}
&R(\beta,L) = f_R(X) +  L^{-\omega} g_R(X) + \ldots \,, \label{scalbeh}\\
&X = (\beta-\beta_c)L^{1/\nu} \,, \label{Xdef}
\end{align}
where $f_R(X)$ is a function that is universal up to a multiplicative
scale of its argument, $\nu$ is the critical exponent of the
correlation length and $\omega$ is the exponent associated with the
leading irrelevant operator.  By expanding Eq.~\eqref{scalbeh} around
$\beta_c$, corresponding to $X=0$, we may write
\begin{eqnarray} \label{rxiansatz}
R(\beta,L)&\approx& R^* + \sum_{k=1}^n a_k X^k 
+ L^{-\omega} \sum_{k=0}^m b_k X^k\,,
\end{eqnarray}
where, as in Eq.~\eqref{scalbeh}, we have neglected next-to-leading
scaling corrections. Using this expression it is possible to estimate
$\beta_c$ and $\nu$ from numerical determinations of $R$.

Since $R_\xi$ defined in Eq.~\eqref{rxidef} is an increasing function of
$\beta$, we may write
\begin{equation}\label{uvsrxi}
U(\beta,L) = F_U(R_\xi) + O(L^{-\omega})\,,
\end{equation}
where $F_U$ now depends on the universality class only, without any
non-universal multiplicative factor. This is true once the boundary
conditions and the shape of the lattice have been fixed, provided one
uses corresponding quantities in the different models, see, e.g.,
Ref.~\cite{PV-19} and the discussion in Sec.~\ref{nf2nc2res}. The
scaling \eqref{uvsrxi} is particularly convenient to test
universality-class predictions, since it permits easy comparisons
between different models without any tuning of nonuniversal
parameters.
 
Finally, we also mention that the susceptibility is expected to scale as
\cite{PV-02}
\begin{align}
\chi(\beta,L) &= 
L^{2-\eta} \Big[ f_\chi(X) + L^{-\omega} g_\chi(X) \Big]
\label{chiansatz}\\
&=
L^{2-\eta} \left[ F_\chi(R_\xi) + O(L^{-\omega}) \right]\,,
\label{chiansatz2}
\end{align}
where $f_\chi$ and $F_\chi$ are universal functions, apart from
trivial multiplicative normalizations and a normalization of the
argument in the case of $f_\chi$.

\section{Numerical results}
\label{sec:numerics}

We now present and discuss the results of Monte Carlo (MC)
simulations. We use an overrelaxation algorithm, consisting of a
combination of heat-bath \cite{Creutz:1980zw, Kennedy:1985nu} and
microcanonical \cite{Creutz:1987xi} updates (with ratio 1:5) for the
gauge fields (implemented \emph{\`a la} Cabibbo-Marinari
\cite{Cabibbo:1982zn} for $N_c>2$) and of a combination of Metropolis
\cite{Metropolis:1953am} and microcanonical updates for the scalar
field.  The Metropolis update was tuned to have an acceptance rate of
approximately 30\%.

\begin{figure}[tbp]
\includegraphics*[scale=\graphicscale]{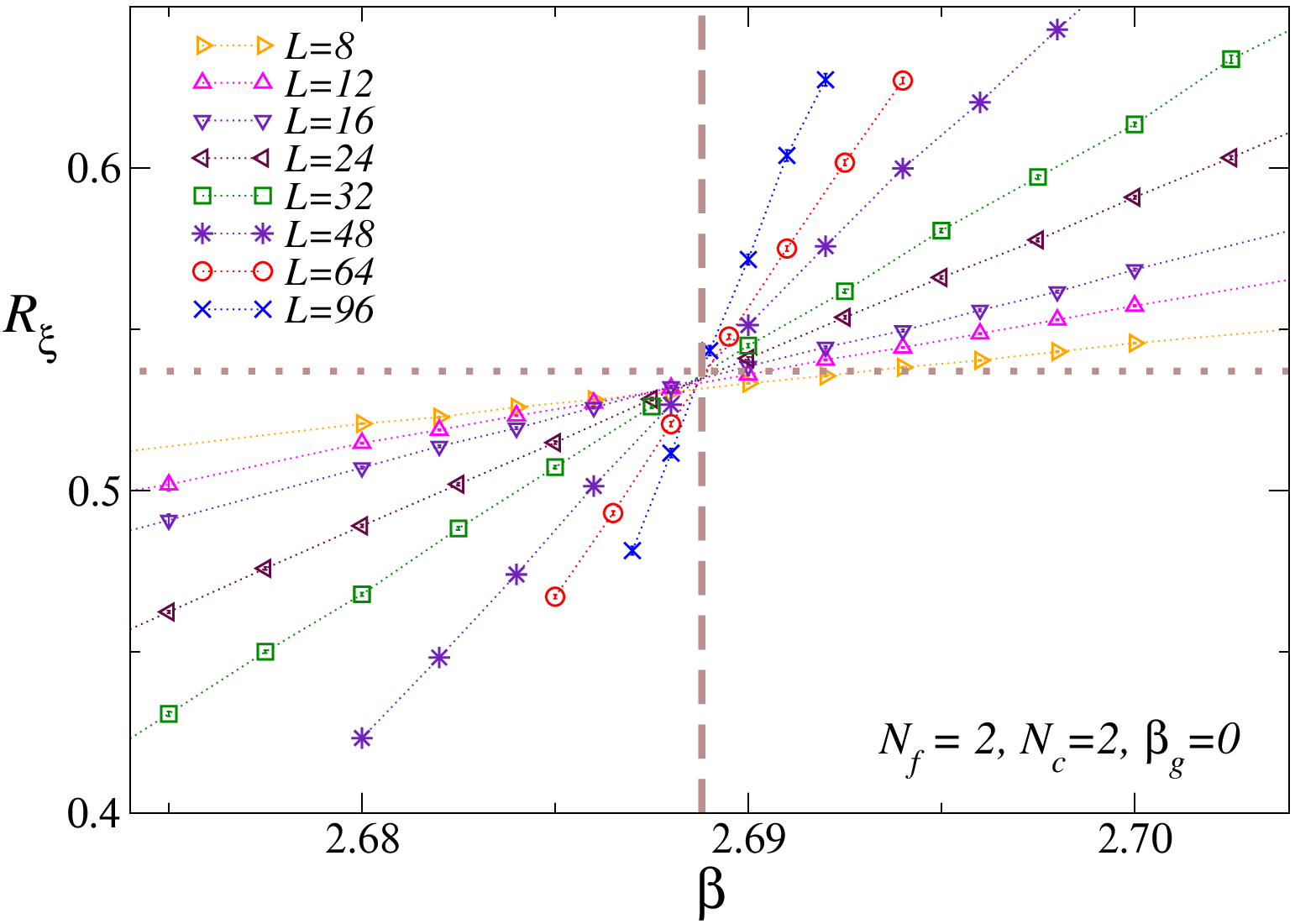}
\caption{$R_\xi$ versus $\beta$ for $N_f=2$, $N_c=2$, and
  $\beta_g=0$. The data for different values of $L$ have a crossing
  point, whose position provides an estimate of the critical point,
  $\beta_c=2.68885(5)$, indicated by the vertical line. The
  horizontal line corresponds to the universal value $R_\xi^*=0.538(1)$ for
   the O(5) vector universality class. }
\label{rxi-f2c2}
\end{figure}

\begin{figure}[tbp]
\includegraphics*[scale=\graphicscale]{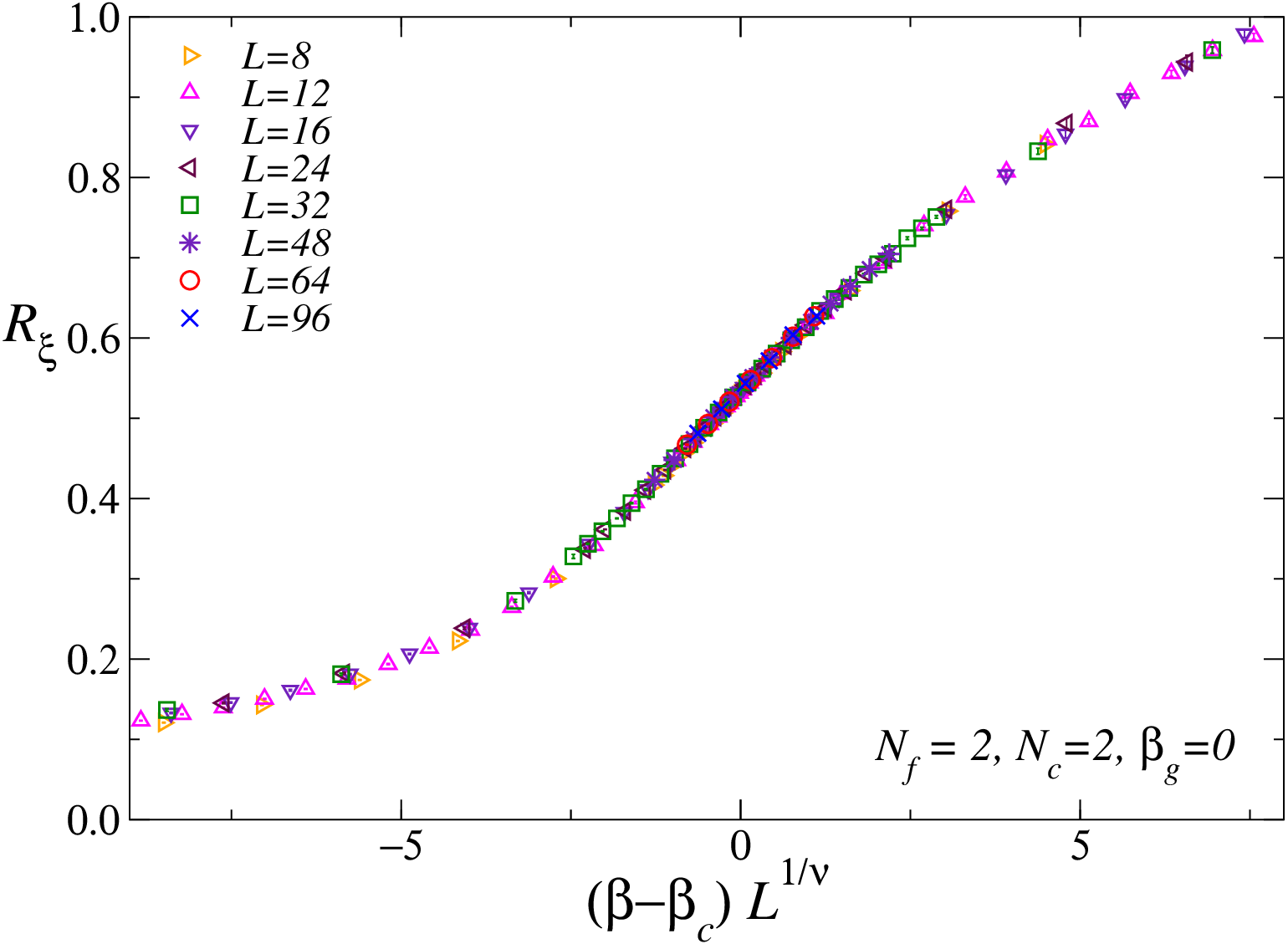}
\caption{$R_\xi$ versus $(\beta-\beta_c)L^{1/\nu}$ for $N_f=2$,
  $N_c=2$, and $\beta_g=0$.  We use $\beta_c=2.68885$ and 
  $\nu=0.779$, the estimate of the correlation-length exponent for
  the O(5) vector
  universality class, see Ref.~\cite{HPV-05}. }  
\label{rxisca-f2c2}
\end{figure}

\begin{figure}[tbp]
\includegraphics*[scale=\graphicscale]{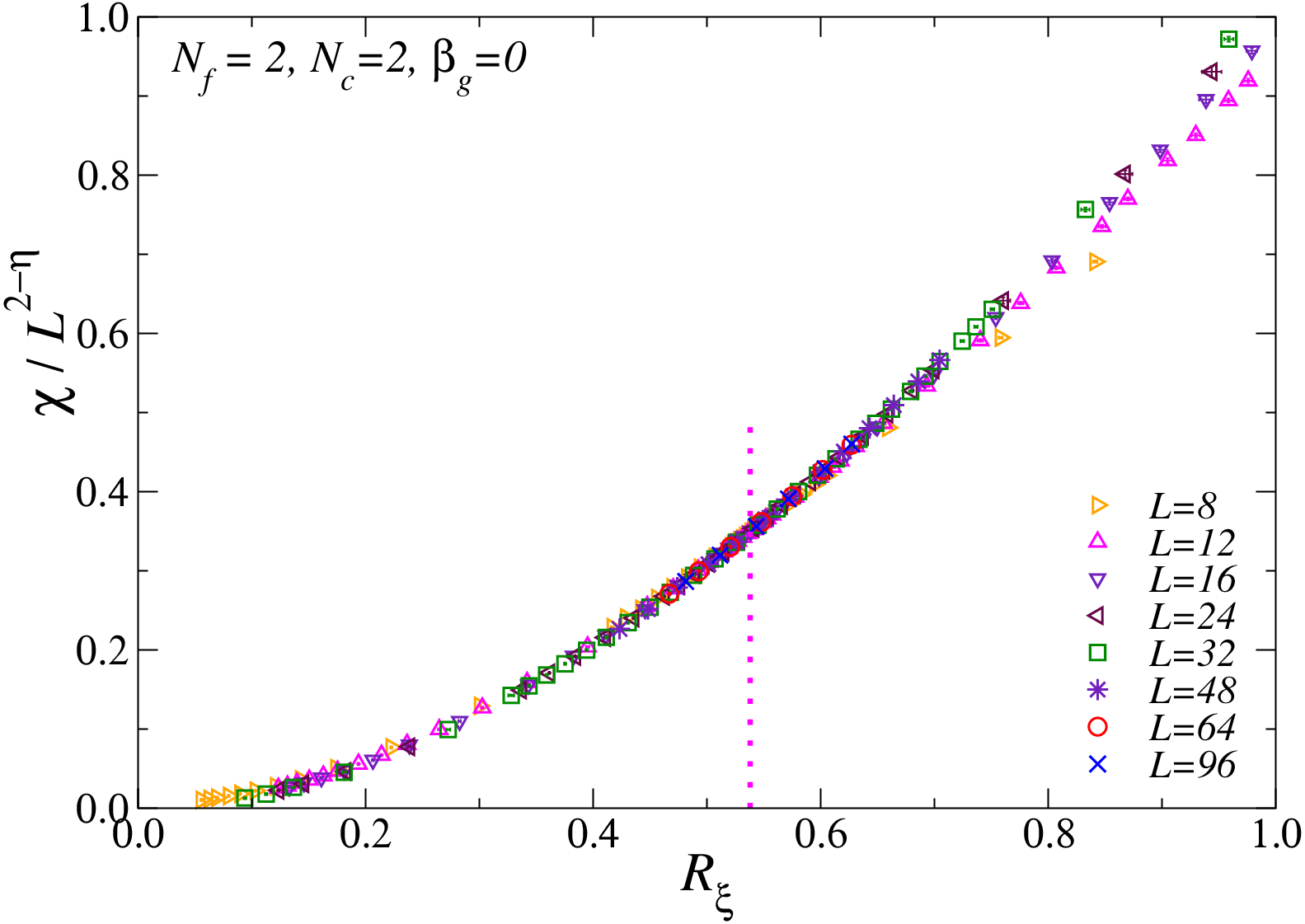}
\caption{Rescaled susceptibility $\chi/L^{2-\eta}$ versus $R_\xi$,
  for $N_f=2$, $N_c=2$, and $\beta_g=0$. We use the 
  estimate $\eta=0.034$, the estimate for the O(5) vector
  universality class, see Ref.~\cite{HPV-05}.  The dotted vertical line
  corresponds to the critical value $R_\xi^*$ for the O(5) vector
  universality class.}
\label{chirxi-f2c2}
\end{figure}

\begin{figure}[tbp]
\includegraphics*[scale=\graphicscale]{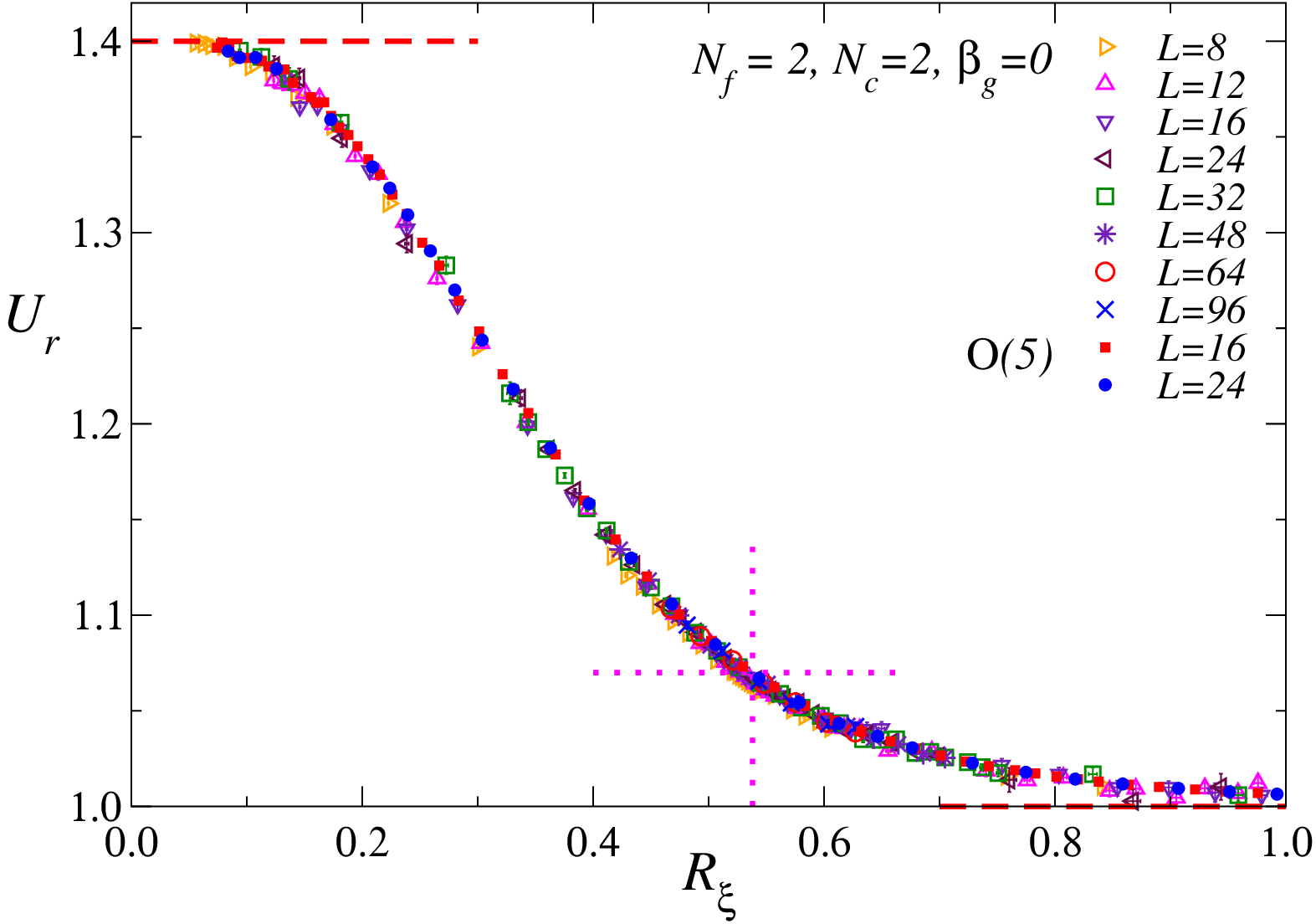}
\includegraphics*[scale=\graphicscale]{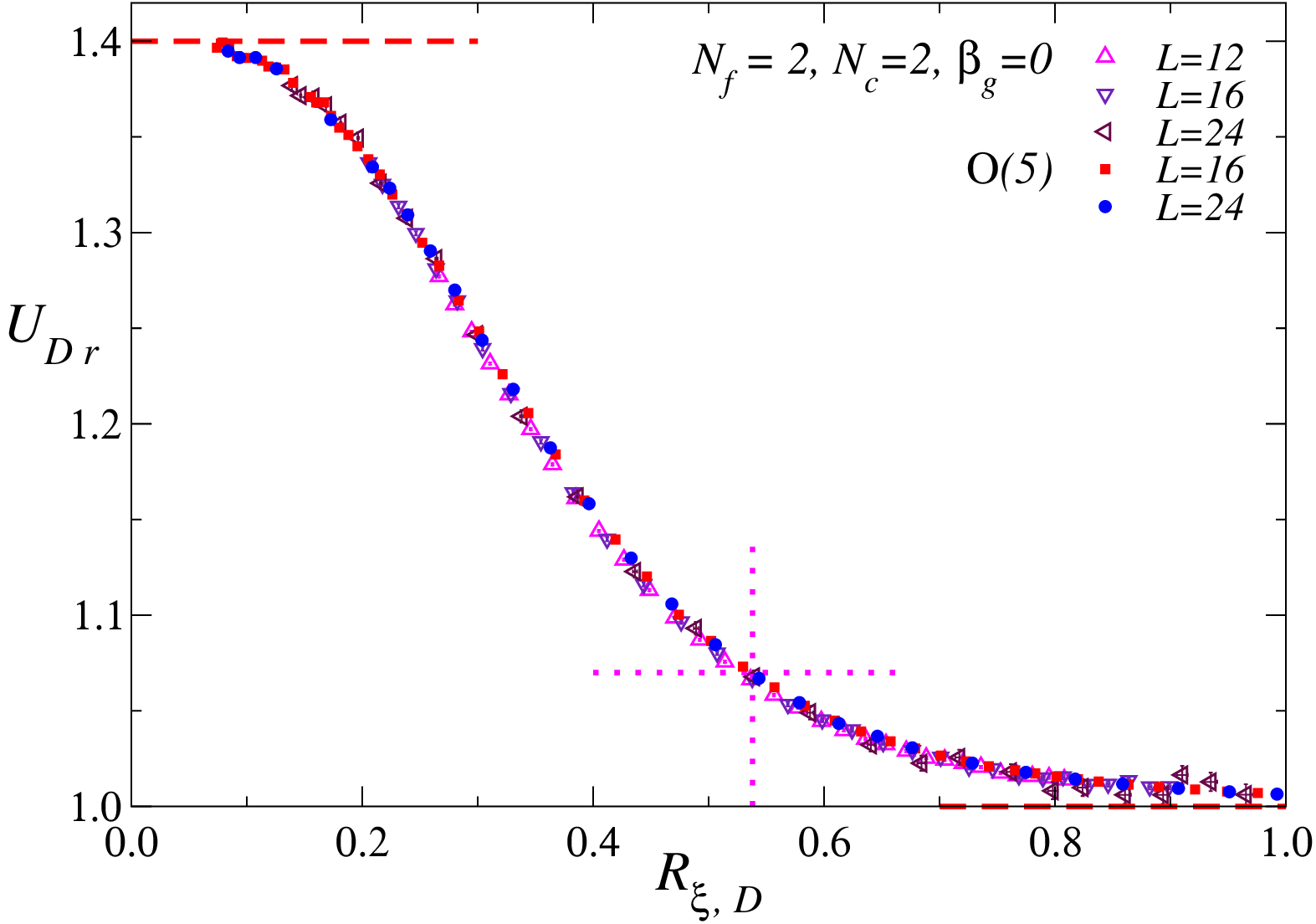}
\caption{Rescaled Binder parameter $U_r$ versus $R_\xi$ (top) and
  rescaled Binder parameter $U_{Dr}$ versus $R_{\xi,D} = \xi_D/L$.
  Results for $N_f=2$, $N_c=2$, and $\beta_g=0$.  Data are in good
  agreement with the the numerical results for the Binder parameter
  obtained by numerical simulations of the O(5) vector lattice model.
  The dotted horizontal and vertical lines correspond to the universal
  values $U^*=1.069(1)$ and $R_\xi^*=0.538(1)$ of the O(5)
  universality class.  The dashed horizontal lines correspond to
  $U_r=7/5$ and $U_r=1$, the asymptotic values for $R_\xi\to 0$ and
  for $R_\xi\to \infty$, respectively. }
\label{urxi-f2c2}
\end{figure}

\subsection{FSS analysis for $N_f=2$ and $N_c=2$}
\label{nf2nc2res}

In this section we present the numerical results obtained for $N_f=2$
and $N_c=2$. We start by analyzing the computationally simplest case
$\beta_g=0$.  In this case we performed simulations on lattices of
size up to $L=96$.
 
In Fig.~\ref{rxi-f2c2} we show the estimates of $R_{\xi}$ for
different values of $L$ and $\beta$. They display the typical behavior
expected at a continuous transition: Different curves have an
approximate crossing point and the slopes increase by increasing the
lattice size. Eq.~\eqref{rxiansatz} can then be used to extract the
critical coupling $\beta_c$ and the critical exponent $\nu$.  For this
purpose we first perform standard nonlinear (unbiased) fits to the
ansatz
\begin{equation}\label{unbfit}
R_\xi=R_\xi^* + a_1 X\,,\qquad X=(\beta-\beta_c) L^{1/\nu}\,,
\end{equation}
using data within the self-consistent window $R_\xi(\beta,L) \in
[R_{\xi}^*(1 - \delta),R_{\xi}^*(1 + \delta)]$.  For $\delta=0.1$ and
$L\ge L_{\rm min}=16$, we obtain $\beta_{c} = 2.68869(2)$, $\nu =
0.775(6)$, and $R_\xi^* = 0.5340(2)$, with $\chi^2/\mathrm{d.o.f.}
\approx 1.5$ (30 data, d.o.f. is the number of degrees of freedom of
the fit). The exponent $\nu$ is consistent with that of the O(5)
vector universality class, whose universal critical exponents
are~\cite{AS-95,HPV-05,FMSTV-05,CPV-03}
\begin{equation}\label{univo5}
\nu=0.779(3)\,,\quad \eta=0.034(1)\,, \quad\omega = 0.79(2)\,.
\end{equation}
\begin{table}
\begin{tabular}{crclcc}
\hline\hline
$\delta$ & 
\multicolumn{1}{c}{$L_{\rm min}$} & 
$\beta_c$  & \multicolumn{1}{c}{$R_{\xi}^*$} & 
$\chi^2/{\mathrm{d.o.f.}}$ & \# data  \\ \hline
0.05     & 8             & 2.68886(3) & 0.5381(3)   & 1.1                        & 45 \\ 
0.10      & 8             & 2.68887(2) & 0.5378(3)   & 1.4                        & 68 \\ 
0.05     & 12            & 2.68880(4) & 0.5372(6)   & 1.1                        & 33 \\ 
0.10      & 12            & 2.68880(3) & 0.5364(5)   & 1.4                        & 52 \\
0.05     & 24            & 2.68886(8) & 0.539(3)    & 1.2                        & 13 \\
0.10      & 24            & 2.68884(6) & 0.538(2)    & 1.3                        & 26 \\   
\hline\hline
\end{tabular}
\caption{Results of the biased fits of $R_\xi$ to the Ansatz
  (\ref{rxiansatz}) with $n=1$, $m=0$, fixing $\nu$ and $\omega$ to
  the O(5) values reported in Eq.~\eqref{univo5}.  Results for
  $N_c=N_f=2$ and $\beta_g = 0$. }\label{tab:bias}
\end{table}
To corroborate this identification, we perform biased fits to
Eq.~\eqref{rxiansatz}, with $n=1$ and $m=0$ (we include a single
scaling correction term), fixing $\nu$ and $\omega$ to the O(5) values
reported in Eq.~\eqref{univo5}.  Again, we use a self-consistent fit
window $R_\xi(\beta,L)\in
[R_\xi^*(1-\delta),R_{\xi}^*(1+\delta)]$. The results are reported in
Table~\ref{tab:bias}.  Our final biased estimates, that take into
account the dependence of the fit parameters on $\delta$ and $L_{\rm
  min}$, are
\begin{equation}
\label{finalest}
\beta_c = 2.68885(5) \,,\quad R_\xi^* = 0.538(2)\,. 
\end{equation}
The errors also take into account the variation of the estimates as
$\nu$ and $\omega$ vary within one error bar.  The corresponding
scaling plot is shown in Fig.~\ref{rxisca-f2c2}, where $R_\xi$ is
plotted versus $X = (\beta-\beta_c)L^{1/\nu}$ using $\beta_c =
2.68885$ and the O(5) value $\nu = 0.779$. The agreement is excellent.
Note also that the estimate of $R_\xi^*$ is consistent with $R_\xi^*=
0.538(1)$, obtained in the O(5) vector model using the vector
correlation function \cite{HPV-05}.  Also the behavior of the
susceptibility $\chi$ is consistent with a transition in the O(5)
universality class. If we fix $\eta$ to the O(5) value [see
  Eq.~\eqref{univo5}], the ratio $\chi/L^{2-\eta}$ scales nicely when
plotted versus $R_\xi$, as expected from the scaling relation
Eq.~(\ref{chiansatz2}), see Fig.~\ref{chirxi-f2c2}.

\begin{figure}[tbp]
\includegraphics*[scale=\graphicscale]{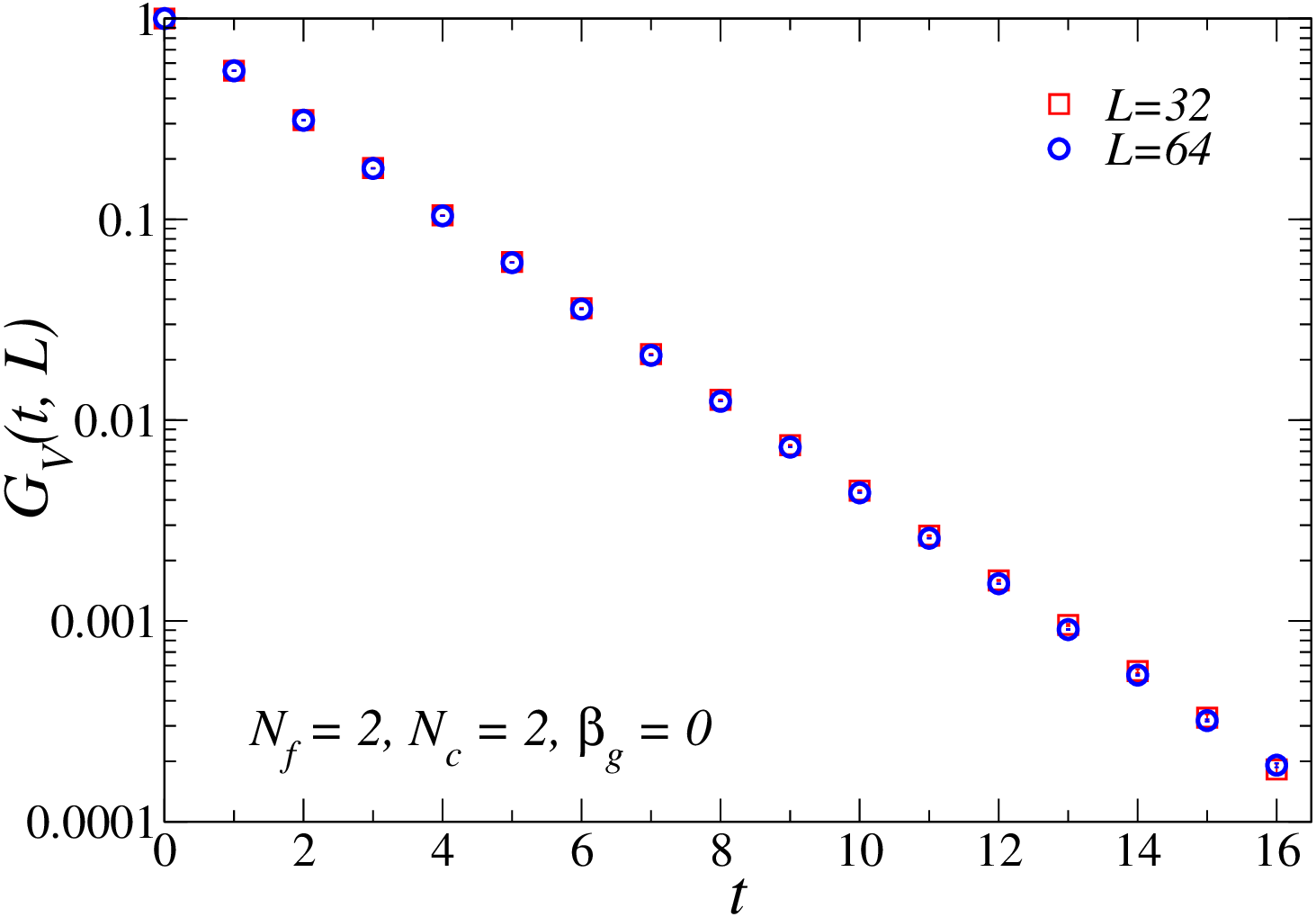}
\caption{The correlation function $G_V$ defined in Eq.~\eqref{Gv}, for
  $N_f=2$, $N_c=2$, and $\beta_g=0$ at $\beta_c$. It shows a
  large-distance exponential behavior $\sim e^{-x/\xi_g}$ with 
  $\xi_g=1.92(2)$.}
\label{gv-f2c2}
\end{figure}

\begin{figure}[tbp]
\includegraphics*[scale=\graphicscale]{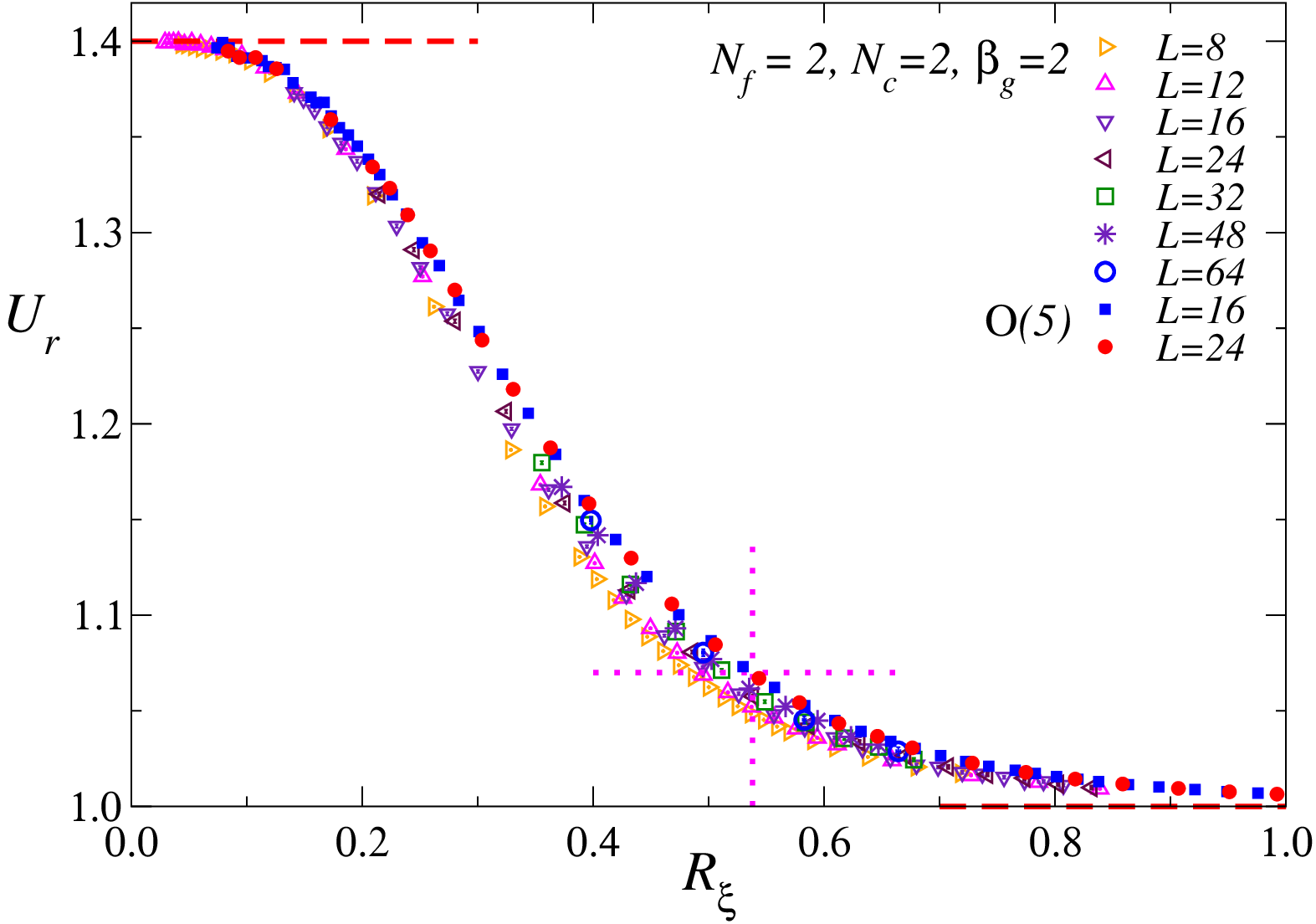}
\caption{Estimates of the rescaled Binder parameter $U_r$ versus
  $R_\xi$, for $N_f=2$, $N_c=2$, and $\beta_g=2$ and of the usual
  Binder parameter for the O(5) vector model. The dotted horizontal
  and vertical lines indicate the universal values $U^*=1.069(1)$ and
  $R_\xi^*=0.538(1)$ of the O(5) universality class.  The dashed
  horizontal lines correspond to the asymptotic values $U_r=7/5$ and
  $U_r = 1$ for $R_\xi\to 0$ and $R_\xi\to \infty$, respectively.  }
\label{urxi-f2c2-bg2}
\end{figure}

Additional evidence that the transition belongs to the O(5) vector
universality class is provided by the analysis of the Binder parameter
$U$ defined in Eq.~(\ref{binderdef}).  To perform the correct
universality check, we should compare corresponding quantities in our
model and in the O(5) vector model. As we discuss at length in
App.~\ref{app:symp}, the Binder parameter that corresponds to the O(5)
parameter is $U_\Gamma$ defined by using ${\cal T}^{\alpha\beta}$, see
Eq.~(\ref{UGammadef}). Using the Sp(2)/O(5) invariance of the theory,
one can easily show that $U_\Gamma$ and $U$ simply differ by a
multiplicative constant, see Eq.~(\ref{app:relationsU}).  Therefore,
the renormalized Binder parameter
\begin{equation}
U_r = {21\over 25} U
\label{uom3}
\end{equation}
should behave as the vector Binder parameter in the O(5) vector
model. If we perform biased fits to Eq.~(\ref{rxiansatz}) analogous to
those we performed for $R_\xi$, we obtain $U_r^*=1.070(1)$, which is
in agreement with the O(5) estimate $U_{{\rm O}(5)}^*=1.069(1)$
reported in Ref.~\cite{HPV-05}.  A conclusive evidence for an O(5)
critical behavior is provided by Fig.~\ref{urxi-f2c2}, where we report
$U_r$ versus $R_\xi$ (upper panel).  The numerical data fall on top of
those obtained in the O(5) vector model.

As we discussed in Sec.~\ref{sec.III}, in the models with $N_c=2$ the
U(1) flavor symmetry breaks at the same $\beta$ where the SU($N_f$) is
broken, since the two groups are subgroups of the larger symmetry
group Sp($N_f$). To verify this point, we have estimated several
observables in terms of the order parameter $D_{\bm x}$ defined in
Eq.~(\ref{Ddefinition}). We have verified that the correlation length
$\xi_D$ defined using the correlation function (\ref{GDdef}) is
identical, within errors, to $\xi$. Moreover, we have studied the
behavior of the Binder parameter $U_D$. Again, to obtain a quantity
that can be directly related to the O(5) Binder parameter, we have
considered, see Eq.~(\ref{app:relationsU}),
\begin{equation}
U_{Dr} = {10\over 7} U_D\,.
\end{equation}
In Fig.~\ref{urxi-f2c2} (lower panel) we report $U_{Dr}$ versus
$R_{\xi,D} = \xi_D/L$.  The data are compared with the O(5)
corresponding data, observing again an excellent agreement.

Finally, we have computed the correlation function $G_V(t)$, defined
in Eq.~\eqref{Gv}.  As it is evident from Fig.~\ref{gv-f2c2}, it is
short-ranged and essentially independent of $L$ even at the critical
point. It has a very clear exponential behavior, $G_V(t)\sim
\exp(-x/\xi_g)$, with $\xi_g=1.92(2)$, independently of the size $L$.
We also analyzed the Polyakov loop which is expected to behave as
$e^{-L/\xi_P}$. The estimates of $\xi_P$ are close to those of
$\xi_g$, but with significantly larger errors.

We have also verified that the analogous results are obtained for
$\beta_g\neq 0$.  For this purpose we performed MC simulations at
$\beta_g=-2$ (using lattices up to $L=32$) and at $\beta_g=2$ (using
lattices up to $L=48$). In both cases data fully support the presence
of a continuous transition in the O(5) universality class. As an
example, in Fig.~\ref{urxi-f2c2-bg2} we plot $U_r$ versus $R_\xi$ for
$\beta_g=2$. Again, the data fall on top of the corrisponding ones
obtained in the O(5) vector model.  Biased fits to
Eq.~(\ref{rxiansatz}) allow us to obtain the estimates
$\beta_c(\beta_g=-2)=3.794(2)$ and
$\beta_c(\beta_g=2)=1.767(1)$. While the critical coupling at
$\beta_g=2$ is significantly lower than the value
$\beta_c(\beta_g=0)\approx 2.689$, it is still quite larger than the
value $\beta_c=0.96339(1)$ which is attained in the limit of large
$\beta_g$, when the model become equivalent to the O(8) vector
model~\cite{DPV-15}.  This could explain the absence of significant
crossover effects in our data induced by the unstable O($2N_fN_c$)
fixed point at $\beta_g\to\infty$, which have instead been observed in
the abelian case \cite{PV-19-2}.  We finally note that the approach to
the asymptotic scaling behavior is significantly slower for
$\beta_g=2$ than for $\beta_g=0$, see Fig.~\ref{urxi-f2c2-bg2}.  This
is likely related to the fact that the gauge length scale $\xi_g$ at
the transition is larger at $\beta_g=2$ than at $\beta_g=0$. Indeed,
we find $\xi_g(\beta_g=2)=2.46(4)$, to be compared with
$\xi_g(\beta_g=0)=1.92(2)$.

The above results provide a robust evidence that the lattice scalar
chromodynamics for $N_f=2$ and $N_c=2$ undergoes a continuous
transition in the O(5) universality class.  This result agrees with
the predictions of the  LGW approach, assuming that the critical behavior is
determined by the global symmetry group and that the gauge degrees of
freedom are irrelevant.

\subsection{FSS analysis for $N_f=2$ and $N_c=3,4$}
\label{nf2nc3res}

\begin{figure}[tbp]
\includegraphics*[scale=\graphicscale]{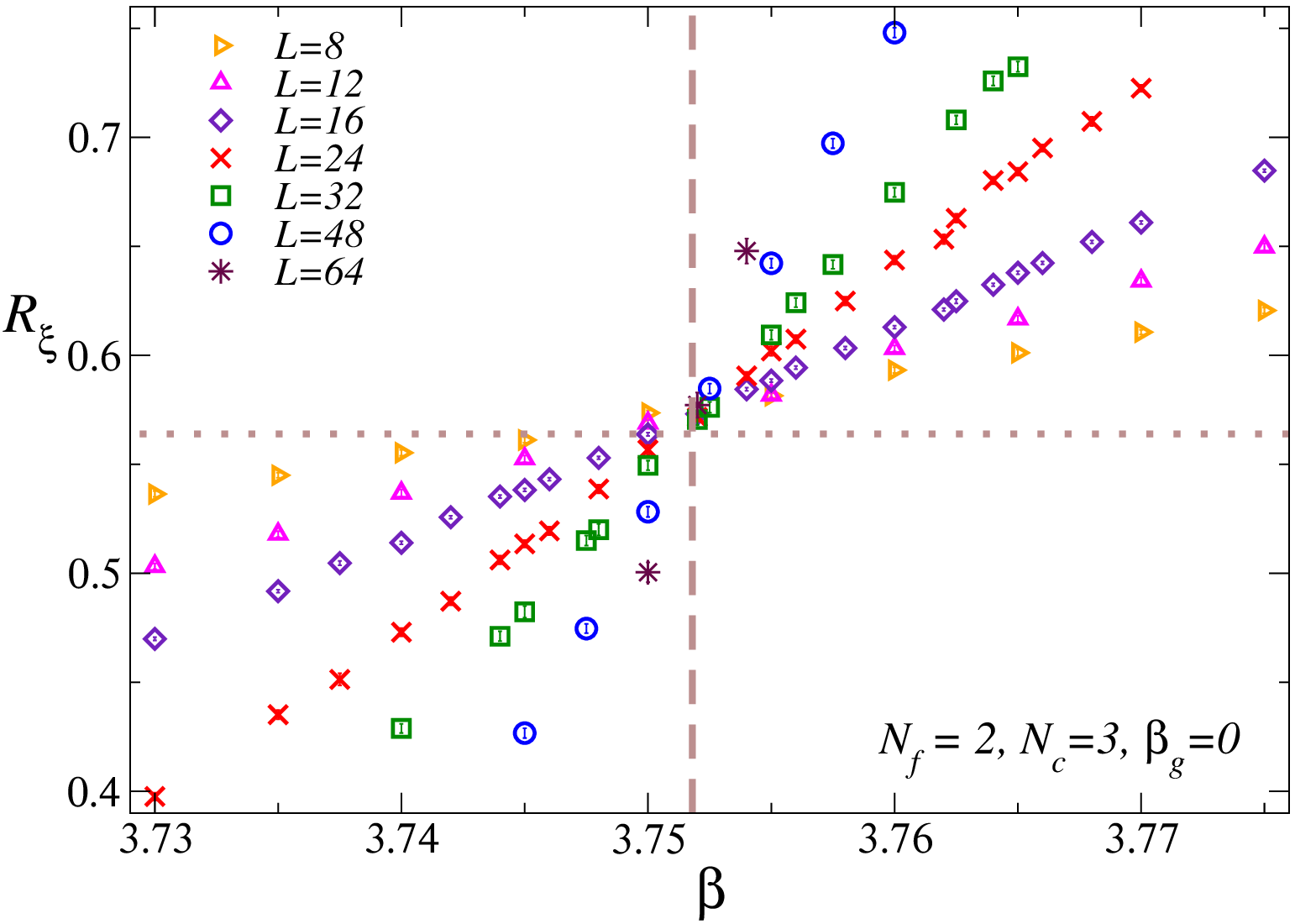}
\caption{ $R_\xi$ versus $\beta$ for $N_f=2$, $N_c=3$, and
  $\beta_g=0$.  The data for different values of $L$ show a crossing
  point, whose position provides an estimate of the critical point,
  $\beta_c=3.7518(2)$, indicated by the vertical line.  The horizontal
  line corresponds to the universal value $R_\xi^*=0.5639(2)$ of the
  O(3) vector universality class.}
\label{rxi-f2c3}
\end{figure}

\begin{figure}[tbp]
\includegraphics*[scale=\graphicscale]{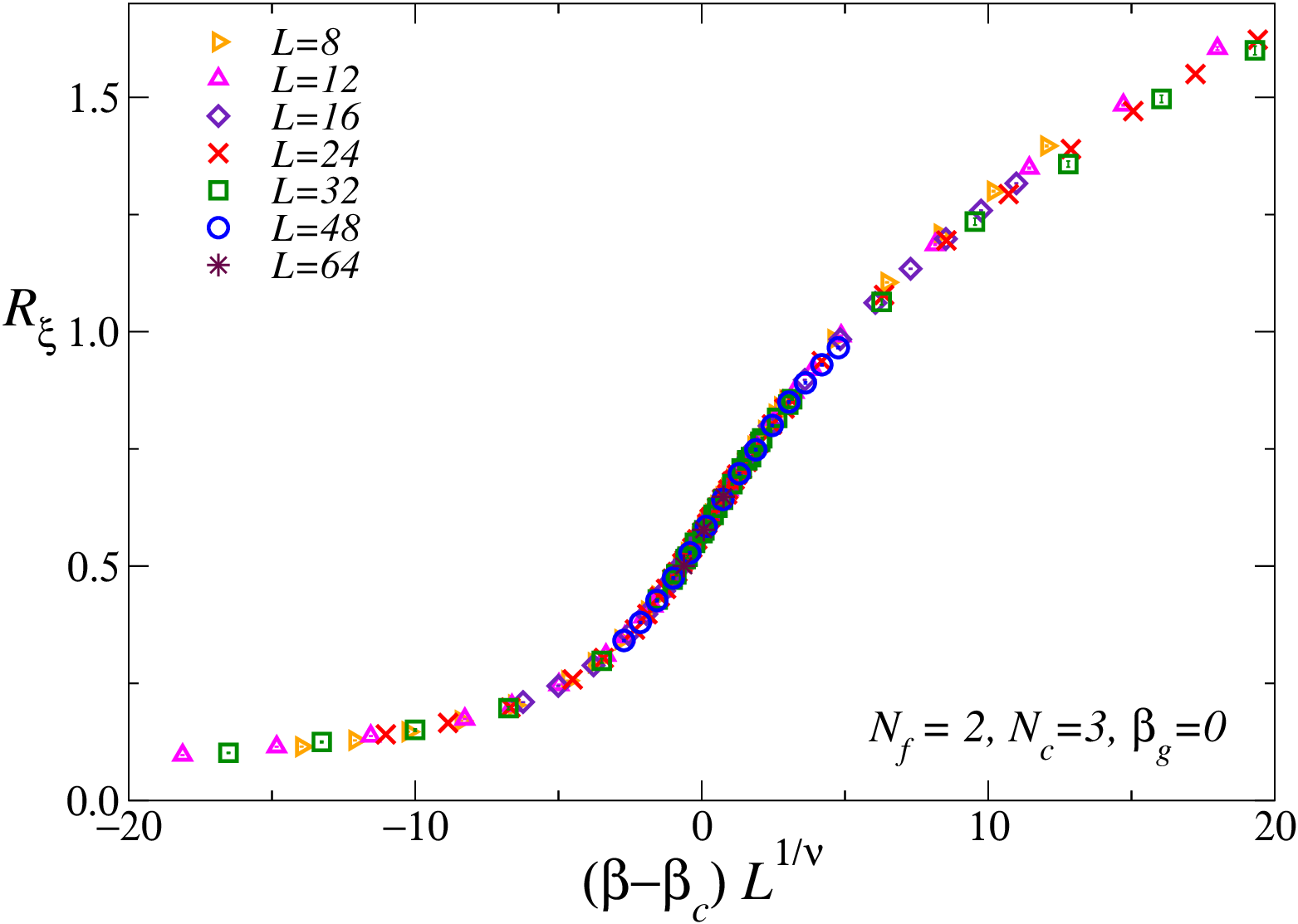}
\caption{ $R_\xi$ versus $(\beta-\beta_c)L^{1/\nu}$ for $N_f=2$,
  $N_c=3$, and $\beta_g=0$.  We use $\beta_c=3.7518$ and
 $\nu=0.7117$, the correlation-length exponent for  the O(3) vector
  universality class. }
\label{rxisca-f2c3}
\end{figure}

In this section we consider the model for $N_f=2$ and $N_c=3,4$.  For
$N_c=3$ and $\beta_g=0$ we heve performed simulation up to $L=64$. In
Fig.~\ref{rxi-f2c3} we report $R_\xi$ as a function of $\beta$.  We
observe a crossing point for $\beta\approx 3.75$.  To determine the
nature of the transition, we again proceed by first performing
standard nonlinear (unbiased) FSS fits of the $R_\xi$ data to the
simplest ansatz Eq.~\eqref{unbfit}, using data within the
self-consistent window $R_\xi(\beta,L) \in [R_\xi^* (1 -\delta),
  R_\xi^*(1 + \delta)]$. For $\delta=0.1$ and $L\ge L_{\rm min}=8$, we
obtain $\beta_{c} = 3.7523(1)$, $\nu = 0.705(10)$, and $R_\xi^* =
0.5771(5)$, with $\chi^2/\mathrm{d.o.f.} \approx 1.4$ (28 data). The
critical exponent $\nu$ is consistent with that of the O(3) vector
universality class, as predicted by the LGW theory. Indeed, the
universal critical exponents and RG invariant quantities of the O(3)
universality class which are relevant for our study
are~\cite{HV-11,CHPRV-02,GZ-98}
\begin{align}
&\nu=0.7117(5)\,,\ \eta = 0.0378(3)\,,\ \omega=0.782(13)\,,\ \label{o3exp}\\
&R_\xi^*=0.5639(2)\,,\ U^*=1.1394(3)\,. \label{rginvo3}
\end{align}

\begin{table}
\begin{tabular}{crllcc}
\hline \hline
$\delta$ & \multicolumn{1}{c}{$L_{\rm min}$} & 
     \multicolumn{1}{c}{$\beta_c$}    & 
      \multicolumn{1}{c}{$R_{\xi}^*$} & 
     $\chi^2/{\mathrm{d.o.f.}}$ & \# data  \\ \hline
0.1      & 8             & 3.75182(9)  & 0.5673(12)  & 1.1                        & 27 \\ 
0.1      & 12            & 3.75186(16) & 0.569(4)    & 1.2                        & 20 \\ 
0.1      & 24            & 3.7521(4)   & 0.577(16)   & 0.8                        & 30 \\ 
0.1      & 32            & 3.7519(11)  & 0.57(6)     & 0.6
& 17 \\
\hline \hline
\end{tabular}
\caption{Results of the biased fits for $R_\xi$ to Eq.~(\ref{rxiansatz}) with 
$n=1$ and $m=2$, fixing $\nu$ and $\omega$ to the O(3) values reported in 
Eq.~(\ref{o3exp}). Results for $N_f=2$, $N_c=3$, and $\beta_g = 0$.
}
\label{tab:biasedfitf2c3}
\end{table}

\begin{figure}[tbp]
\includegraphics*[scale=\graphicscale]{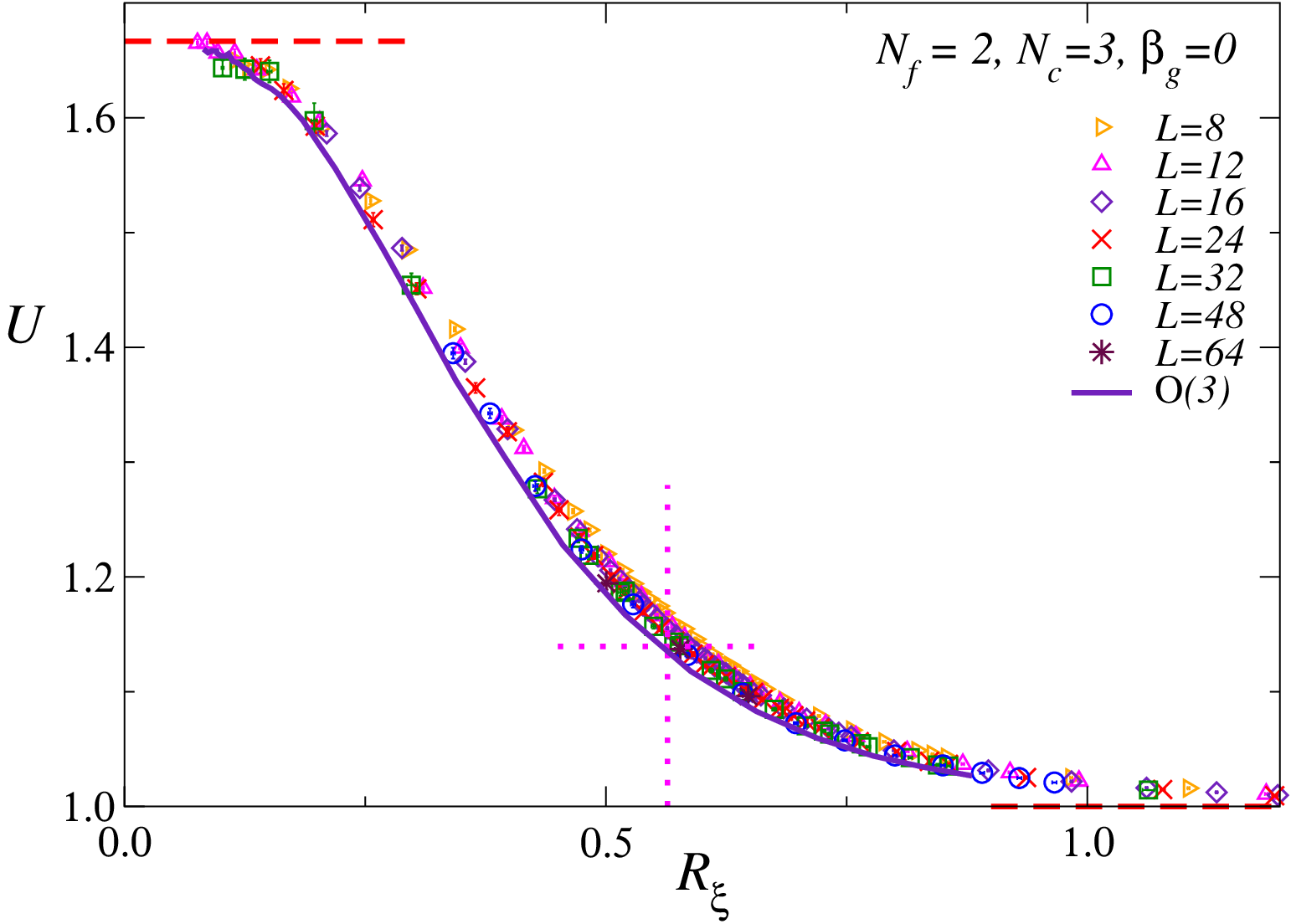}
\caption{ The Binder parameter $U$ versus $R_\xi$, for $N_f=2$,
  $N_c=3$, and $\beta_g=0$.  The data clearly converge to the O(3)
  vector universal curve (continuous curve).  The dotted horizontal
  and vertical lines correspond to the universal values
  $U^*=1.1394(3)$ and $R_\xi^*=0.5639(2)$ of the O(3) universality
  class.  The dashed horizontal lines correspond to $U=5/3$ and $U=1$,
  the asymptotic values of $R_\xi\to 0$ and for $R_\xi\to \infty$,
  respectively.  }
\label{urxi-f2c3}
\end{figure}

\begin{figure}[tbp]
\includegraphics*[scale=\graphicscale]{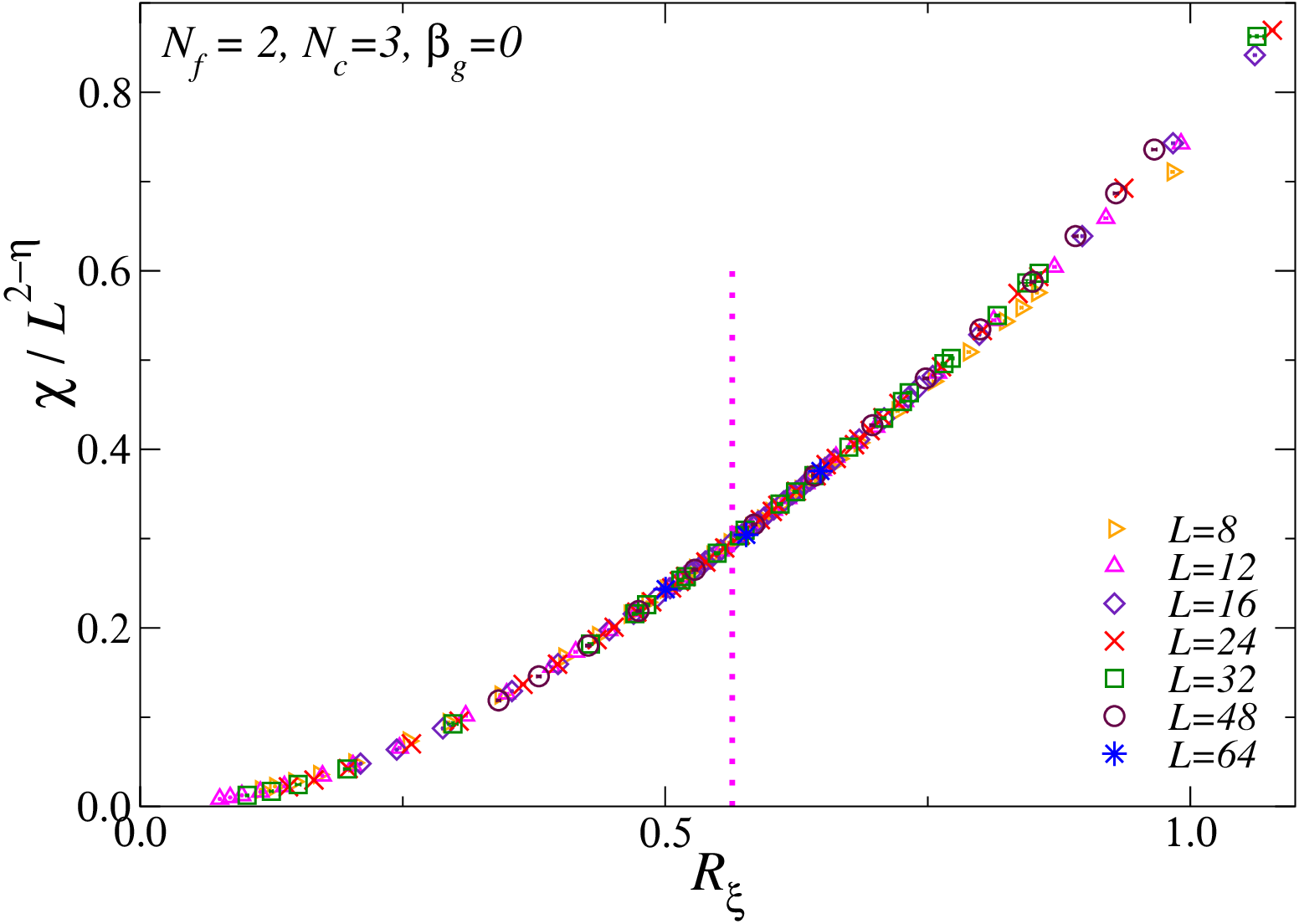}
\caption{The rescaled susceptibility $\chi/L^{2-\eta}$ with
  $\eta=0.0378$, the exponent value in the O(3) vector universality
  class, versus $R_\xi$, for $N_f=2$, $N_c=3$, and $\beta_g=0$.  The
  dotted vertical line corresponds to $R_\xi^*$.}
\label{chirxi-f2c3}
\end{figure}

Additional evidence for an O(3) critical behavior is obtained by
performing biased fits to Eq.~\eqref{rxiansatz} with $n=1$ and $m=0$,
fixing $\nu$ and $\omega$ to the O(3) values reported in
Eq.~(\ref{o3exp}). As before, we use data within the self-consistent
window $R_\xi(\beta,L) \in [R_\xi^* (1 -\delta), R_\xi^*(1 +
  \delta)]$. The results are reported in
Table~\ref{tab:biasedfitf2c3}. The estimates of $R_\xi^*$ are nicely
consistent with the O(3) estimate $R_\xi^*=0.5639(2)$.  A similar
analysis can be done using the Binder parameter $U$.  Using $L_{\rm
  min}=8$, we obtain the estimates $\beta_c = 3.7519(2)$ and $U^* =
1.139(3)$, with $\chi^2/\mathrm{d.o.f}\approx 1.2$ (27 data). Again,
the estimate of $U^*$ is in good agreement with the O(3) value
$U^*=1.1394(3)$.  Our final estimate of the critical temperature,
obtained by considering the various systematic errors, is
\begin{equation}\label{betaco3}
\beta_c=3.7518(2)\,.
\end{equation}  
In Figs. \ref{rxisca-f2c3}, \ref{urxi-f2c3}, and \ref{chirxi-f2c3} we
show different scaling plots that clearly confirm that the transition
belongs to the O(3) universality class. In particular, the data of $U$
plotted versus $R_\xi$, see Fig.~\ref{urxi-f2c3}, are nicely
consistent with the results obtained in numerical simulations of the
O(3) vector model.

\begin{figure}[tbp]
\includegraphics*[scale=\graphicscale]{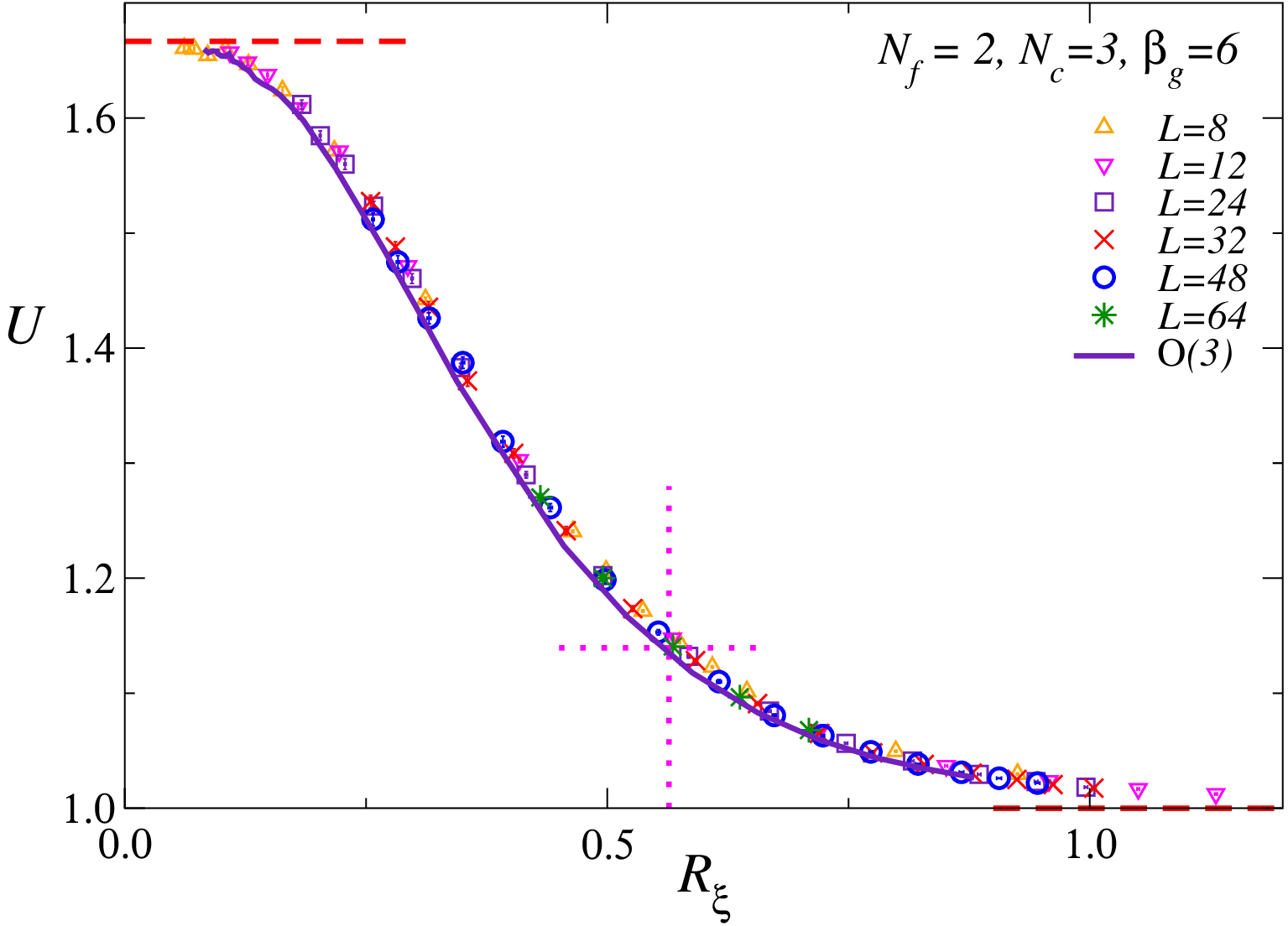}
\includegraphics*[scale=\graphicscale]{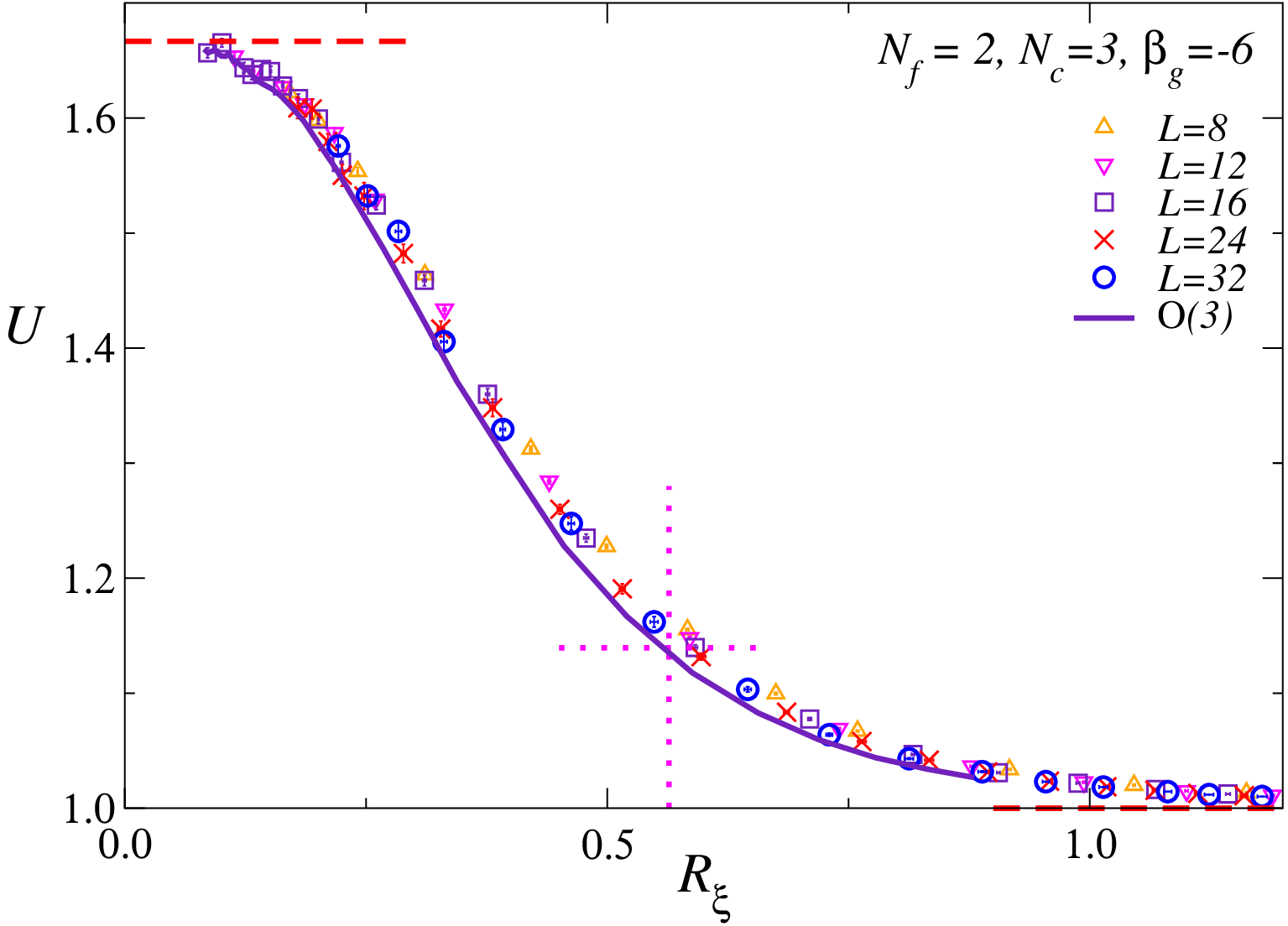}
\caption{The Binder parameter $U$ versus $R_\xi$, for $N_f=2$,
  $N_c=3$. In the lower panel we report results for $\beta_g=-6$ up to
  $L=32$, in the upper panel results for $\beta_g=6$ up to $L=64$.
  The data appear to converge to the O(3) vector universal curve
  (continuous line). The dotted horizontal and vertical lines
  correspond to the universal values $U^*=1.1394(3)$ and
  $R_\xi^*=0.5639(2)$ of the O(3) universality class.  The dashed
  horizontal lines correspond to $U=5/3$ and $U=1$, the asymptotic
  values for $R_\xi\to 0$ and $R_\xi\to \infty$.  }
\label{urxi-f2c3-bg}
\end{figure}

\begin{figure}[tbp]
\includegraphics*[scale=\graphicscale]{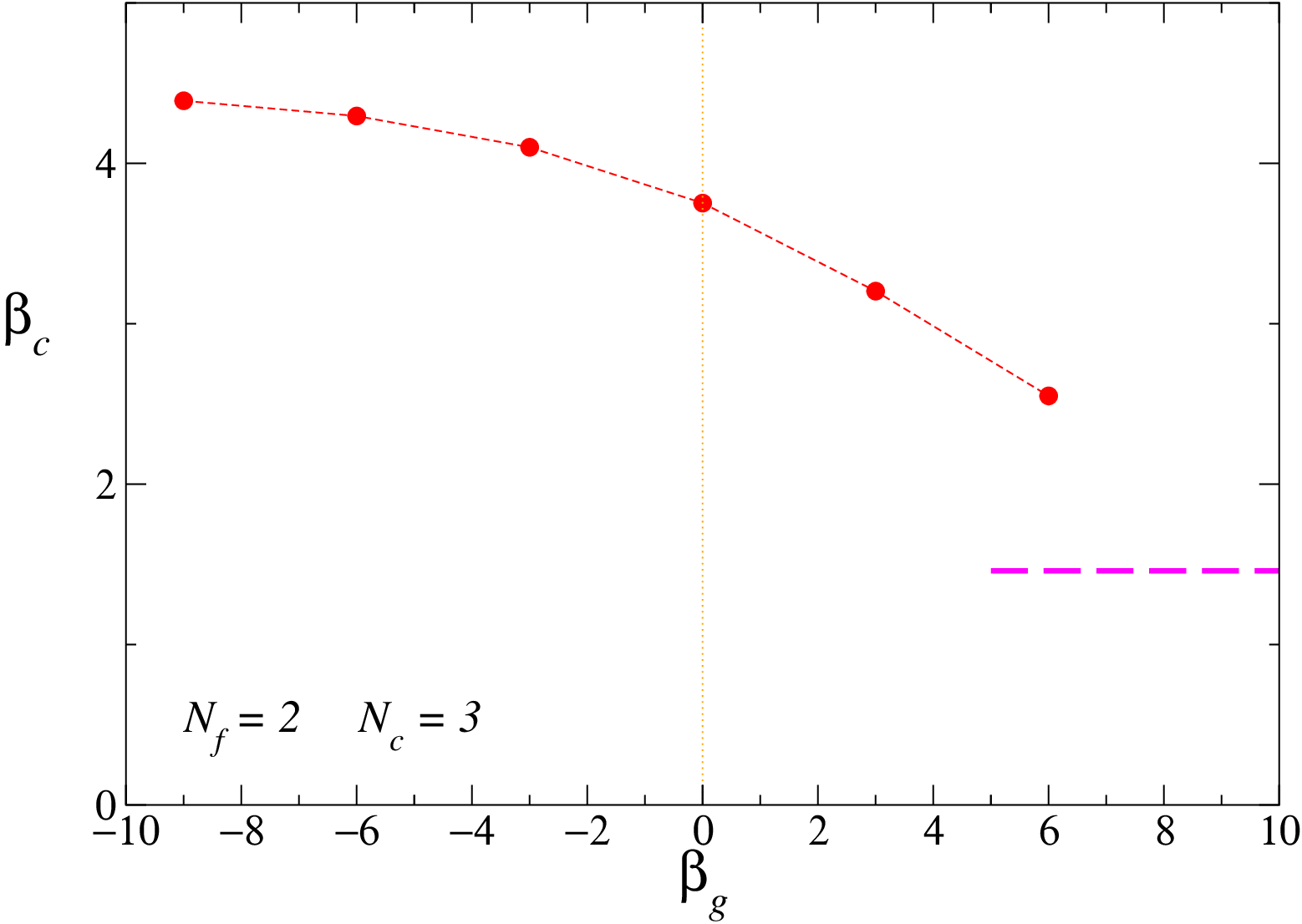}
\caption{Estimates of $\beta_c$ versus $\beta_g$ for the model with
  $N_f=2$, $N_c=3$. The dashed line indicates the  critical value 
  in the
  limit $\beta_g\to\infty$, corresponding to the critical point of the O(12)
  theory, $\beta_c \approx 1.46$, obtained using the results reported
  in Ref.~\cite{CPRV-96}.  The dotted line connecting the data is drawn to
  guide the eyes.
  }
\label{betac-f2c3}
\end{figure}

\begin{figure}[tbp]
\includegraphics*[scale=\graphicscale]{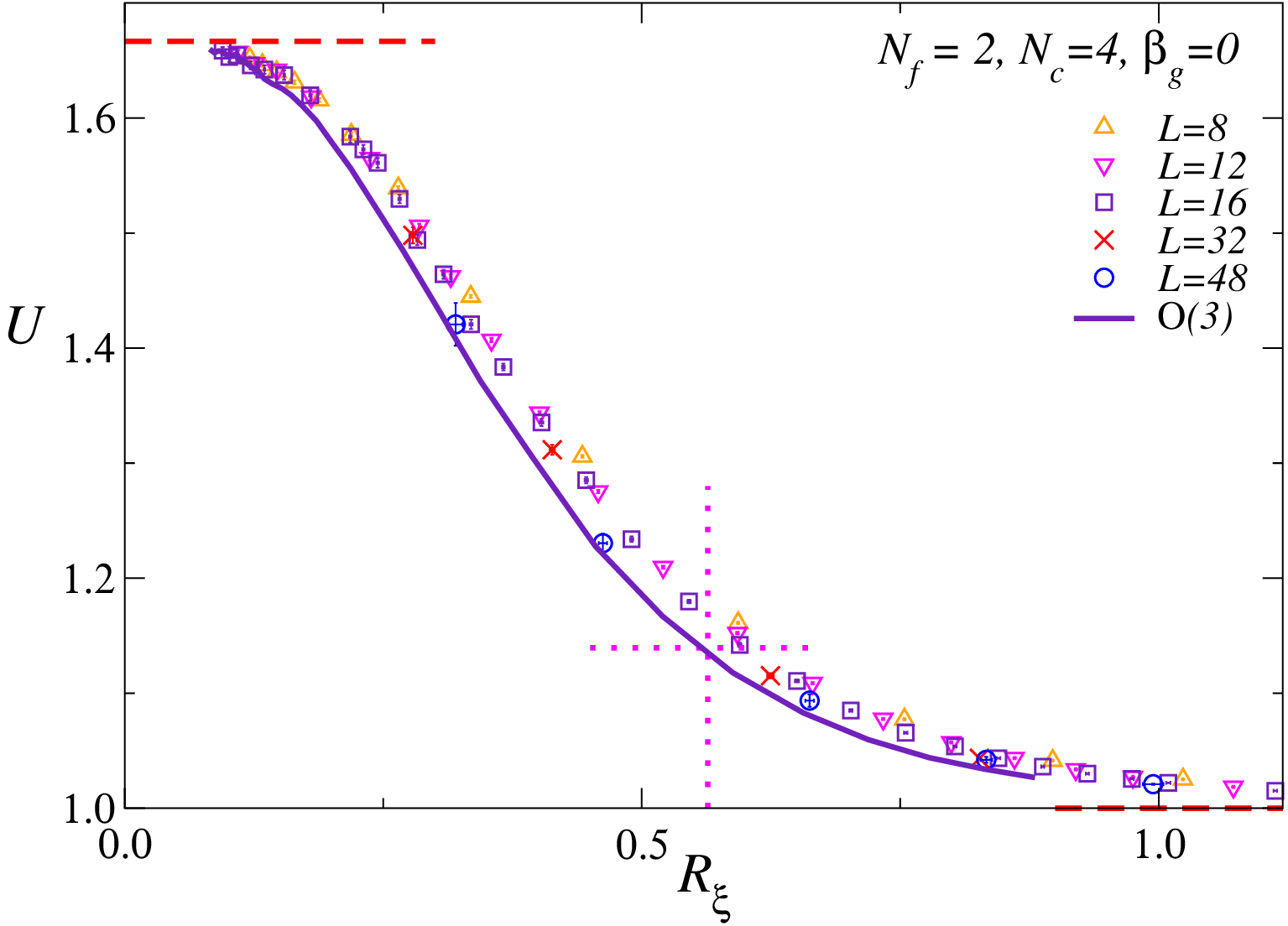}
\caption{The Binder parameter $U$ versus $R_\xi$, for $N_f=2$,
  $N_c=4$, and $\beta_g=0$.  The data appear to converge to the O(3)
  vector universal curve (continuous line).  The dotted horizontal and
  vertical lines correspond to the universal values $U^*=1.1394(3)$
  and $R_\xi^*=0.5639(2)$ of the O(3) universality class.  The dashed
  horizontal lines correspond to $U=5/3$ and $U=1$, the asymptotic
  values for $R_\xi\to 0$ and $R_\xi\to \infty$.  }
\label{urxi-f2c4}
\end{figure}

As in the two color case, we have also checked that the above results
extend to nonvanishing values of $\beta_g$. In particular, simulations
have been performed for a few values of $\beta_g$ between $-9$ and
6. In all cases, the FSS behavior of $U$ as a function of $R_\xi$
supports the O(3) nature of the transition, as can be seen in
Fig.~\ref{urxi-f2c3-bg}, where we report the results for $\beta_g=-6$
and $\beta_g=6$.  Again these results are far from trivial, since the
critical coupling $\beta_c(\beta_g)$ changes from approximately $4.39$
to $2.55$ as we vary $\beta_g$ in the interval $[-9,6]$ (see
Fig.~\ref{betac-f2c3}).  Therefore, the effect of $\beta_g$ on the
dynamics of the system is large.  Also the average gauge energy $E_g$
at criticality changes significantly.  It varies approximately from
$-0.23$ to $0.51$. These values are however still far from the
asymptotic values $\pm 1$ at $\beta_g\to\pm \infty$ and this could
explain the absence of sizable crossover effects in our data. This is
also consistent with the fact that the correlation length associated
with the gauge modes increases with increasing $\beta_g$, but
nevertheless stays quite small: at the transition we obtain
$\xi_g(\beta_g=0)=1.60(2)$, $\xi_g(\beta_g=3)=1.70(2)$, and
$\xi_g(\beta_g=6)=2.02(2)$.

As a final check that, for $N_f=2$ and any $N_c\ge 3$, the transition
always belongs to the O(3) universality class, we performed
MC simulations for $N_c=4$ and $\beta_g=0$. Also in this case the
data of $U$ plotted versus
$R_\xi$  (we have results for $L\le 48$)
clearly approach the O(3) curve as $L$ is increased, as it
can be seen in Fig.~\eqref{urxi-f2c4}. Again, the results confirm the
LGW predictions.

\subsection{FSS analysis for $N_f=3$}
\label{nf3res}

\begin{figure}[tbp]
\includegraphics*[scale=\graphicscale]{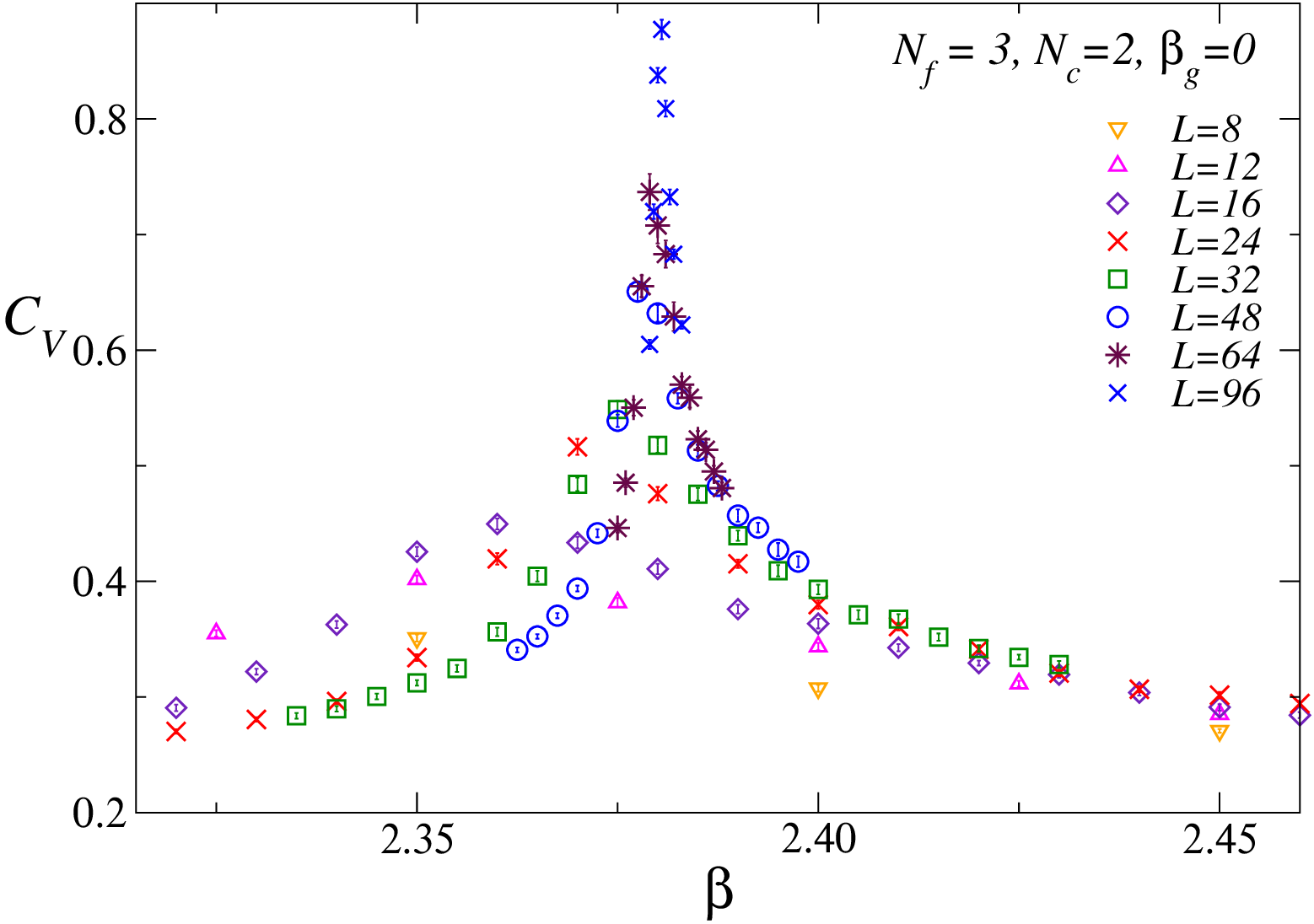}
\caption{The specific heat defined in Eq.~(\ref{ecvdef}) versus
  $\beta$ for $N_c=2$, $N_f=3$, and $\beta_g = 0$.}
\label{cv-f3c2}
\end{figure}

For $N_f=3$ the LGW effective field theory predicts a first-order
phase transition for any number of colors. To verify the prediction,
we perform sumulations for $N_c=2$ and $N_c=3$, fixing always $\beta_g
= 0$.

A standard technique to identify first-order phase transitions
consists in checking if the maximum value of the susceptibility or of
the specific heat scales linearly with the volume. However, for weak
first order transitions such a technique is, in practice, quite often
ineffective: The values of $L$ at which such a behavior sets in are
far larger than those at which simulations can be performed.  This is
indeed what happens, as we discuss below, for $N_c=2$ and 3.

In Fig.~\ref{cv-f3c2} we report the specific heat $C_V$ defined in
Eq.~\eqref{ecvdef} for $N_c = 2$.  It is clear that the specific heat
is apparently diverging as $L$ increases. This allows us to conclude
that the transition, if continuous, does not belong to a universality
class characterized by a negative value of the critical exponent
$\alpha$, like, e.g., the standard O($M$) universality classes
for any $M\ge 2$~\cite{PV-02}.

\begin{figure}[tbp]
\includegraphics*[scale=\graphicscale]{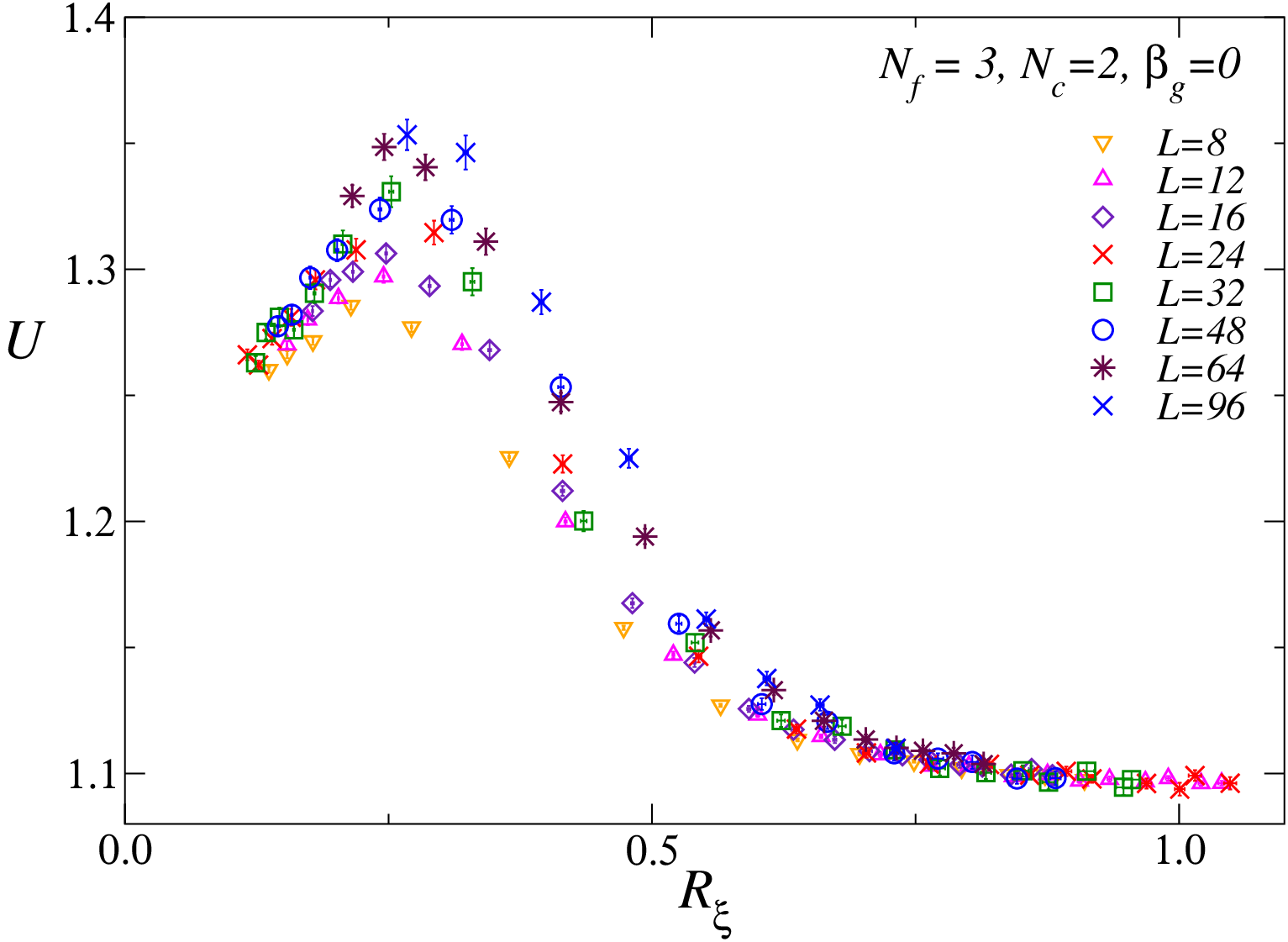}
\includegraphics*[scale=\graphicscale]{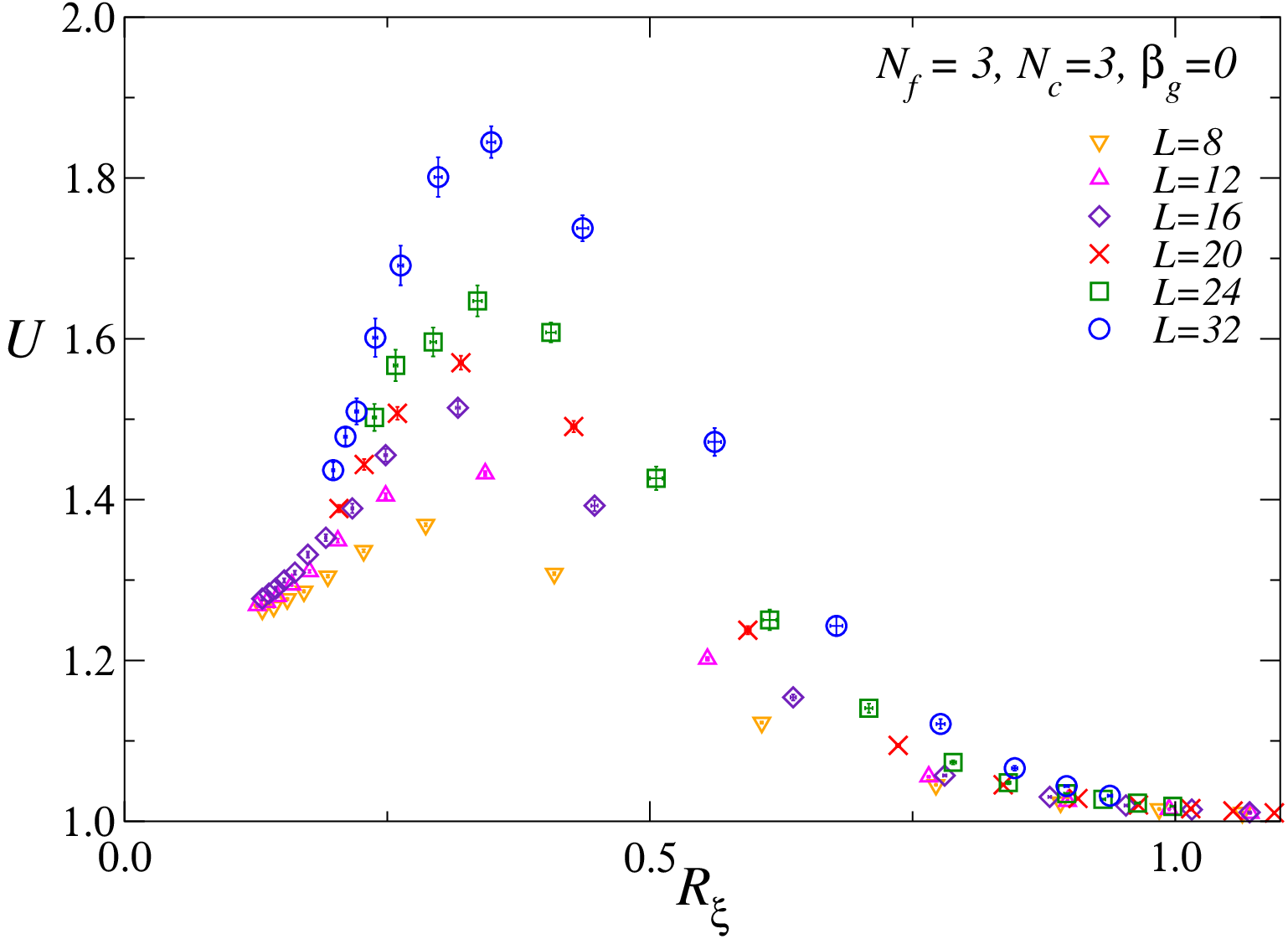}
\caption{The Binder parameter $U$ versus $R_\xi$, for $N_f=3$, $N_c=2$
  (top) and $N_c=3$ (bottom), and $\beta_g=0$.  The presence of a
  maximum of $U$ diverging in the large-$L$ limit is a peculiar
  feature of the behavior at first-order transitions, see, e.g.,
  Refs.~\cite{CLB-86,VRSB-93,PV-19}.}
\label{urxi-f3c2e3}
\end{figure}

\begin{figure}[tbp]
\includegraphics*[scale=\graphicscale]{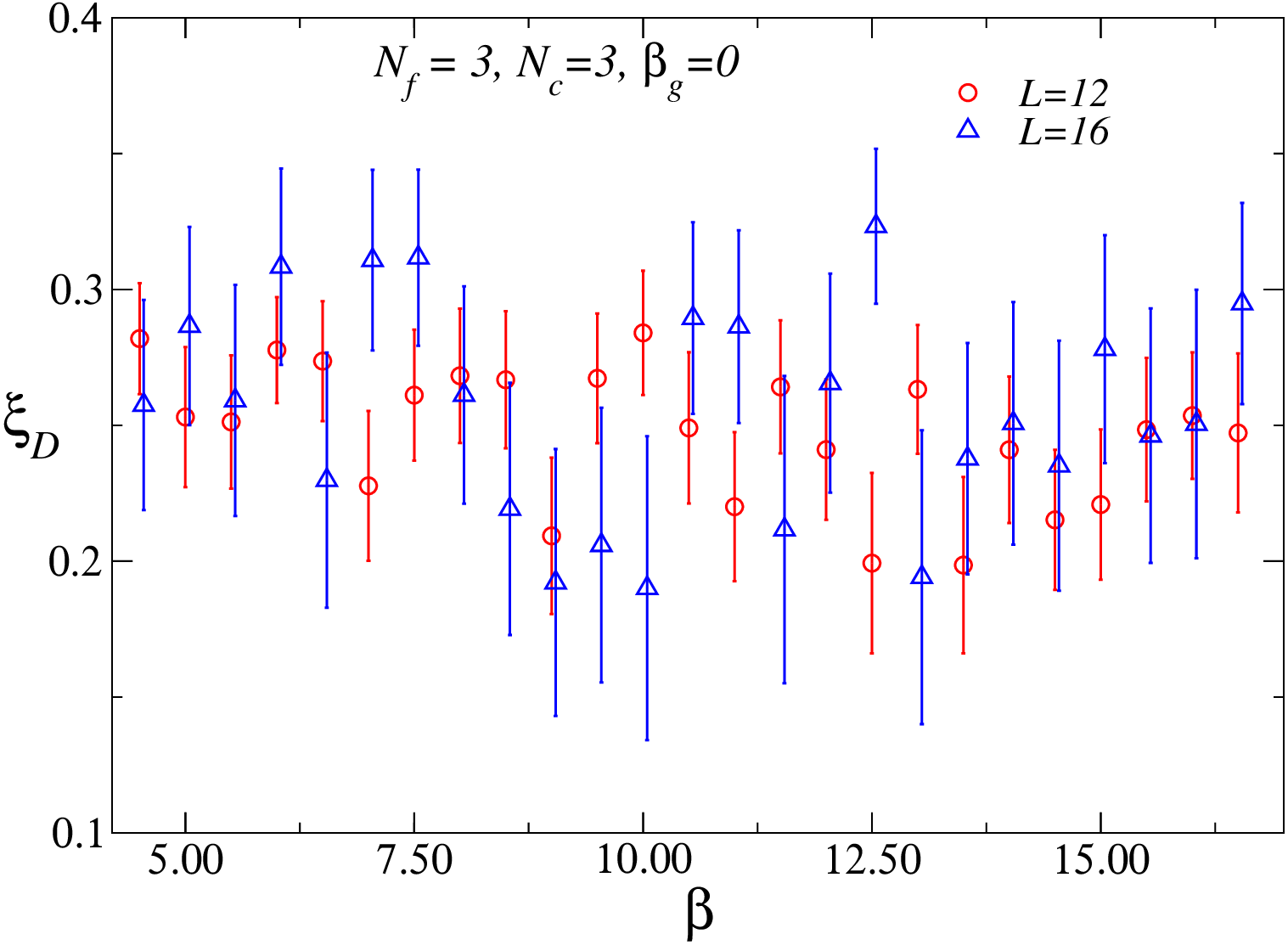}
\includegraphics*[scale=\graphicscale]{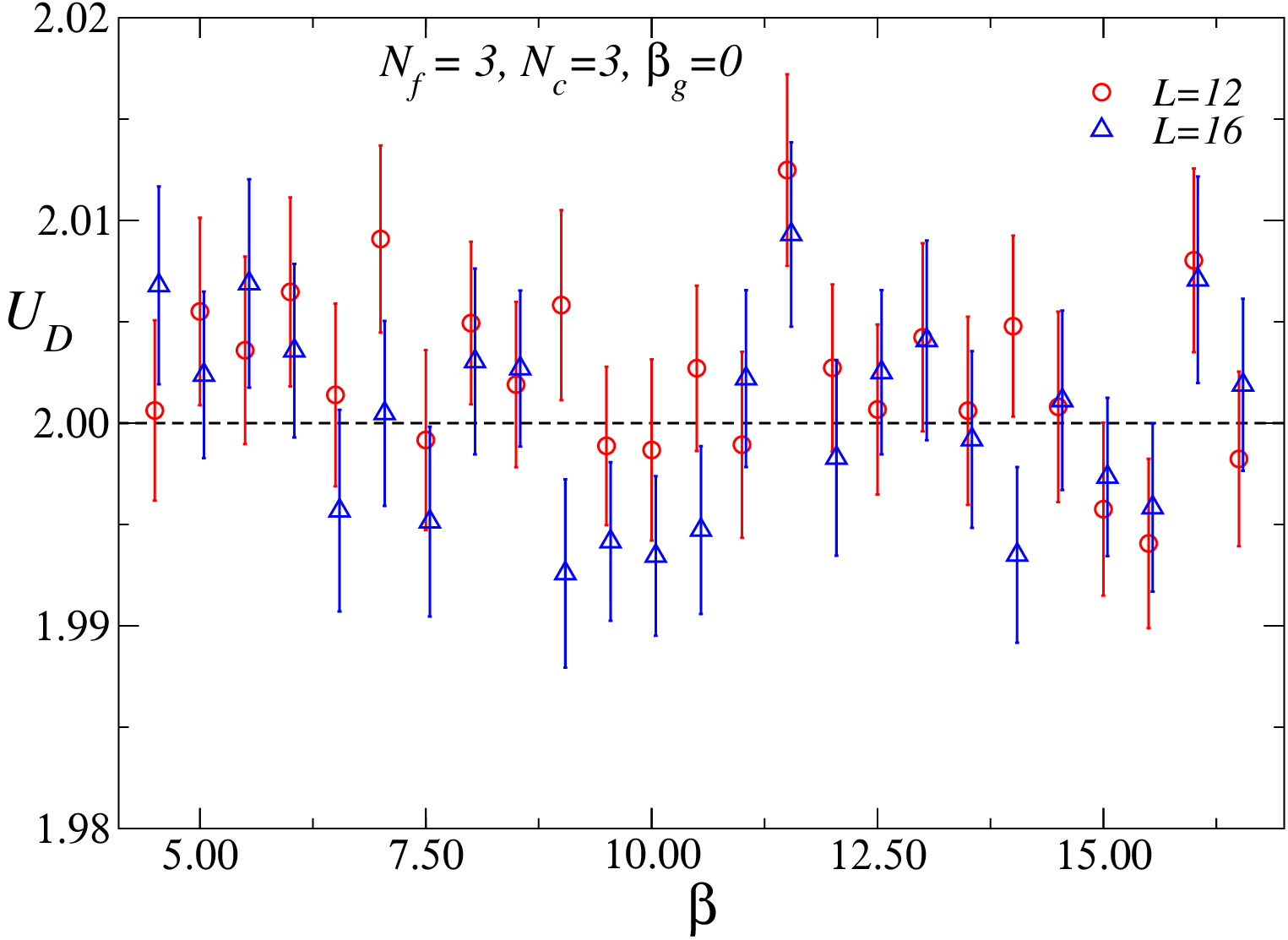}
\caption{The correlation length $\xi_D$ and the Binder parameter 
$U_D$ versus $\beta$,
for $N_f=3$, $N_c=3$, and $\beta_g=0$. We report results for $\beta$ in the range [4,17].
The first-order transition occurs at $\beta_c \approx 3.415$.
}
\label{u1-f3c3}
\end{figure}

In the case of weak first-order transitions, a more useful quantity is
the Binder parameter $U$. At a first-order transition, the maximum
$U_{\rm max}$ of $U$ behaves as~\cite{CLB-86,VRSB-93}
\begin{equation}\label{Ufirst}
U_{\rm max}= a V[1 + O(V^{-1})]\,.
\end{equation}
On the other hand, at a continuous phase transition, $U$ is bounded as
$L\to \infty$. At the critical point $U$ converges to a universal
value $U^*$, while the data of $U$ corresponding to different values
of $R_{\xi}$ collapse onto a common scaling curve as the volume is
increased.  Therefore, $U$ has a qualitatively different scaling
behavior for first- and second-order transitions. In practice, a
first-order transition can be simply identified by verifying that
$U_{\rm max}$ increases with $L$, without the need of explicitly
observing the linear behavior in the volume. A second indication of a
first-order transition is provided by the plot of $U$ versus
$R_{\xi}$. The absence of a data collapse is an early indication of
the first-order nature of the transition, as already advocated in
Ref.~\cite{PV-19}.  In Fig.~\ref{urxi-f3c2e3} we plot the Binder
parameter $U$ versus $R_{\xi}$, for $N_c=2$ and $N_c=3$, respectivele,
and $\beta_g=0$.  In neither of the two cases an acceptable collapse
is obtained and the data display a pronounced peak whose height
increases with increasing volume. We take the absence of scaling as an
evidence  that the transition is not continuous, thus that 
it is of first order in both cases.

We have also investigated the behavior of the observables related to
the breaking of the U(1) flavor symmetry. In Fig.~\ref{u1-f3c3} we
report the correlation length $\xi_D$ and the Binder parameter $U_D$,
defined in Sec.~\ref{sec4.1}. Our numerical results show that the correlation
length $\xi_D$ is always small, even at the transition point $\beta
\approx 3.415$, a clear indication that the U(1) flavor symmetry does
not break. The results for the Binder parameter are completely
consistent: $U_D$ is always compatible with the high-temperature value
$U_D = 2$.

\section{Conclusions}
\label{sec:concl}

In this work we have studied the finite-temperature critical behavior of
the lattice multiflavor chromodynamics model defined by the action,
Eq.~\eqref{hgauge}.  This model is characterized by the presence of a
SU($N_c$) gauge symmetry and of a U($N_f$) or Sp($N_f$) global
symmetry, depending whether $N_c\ge 3$ or $N_c=2$. In all cases, we
find that the system undergoes a finite-temperature phase transition
associated with the condensation of a gauge-invariant order parameter:
the operator $Q_{\bm x}^{ab}$ for $N_c\ge 3$ and the operator ${\cal
  T}_{\bm x}^{\alpha\beta}$ for $N_c=2$.  At the phase transition the global
symmetry SU($N_f$) or Sp($N_f$) is spontaneously broken.

To investigate the possible influence of the gauge degrees of freedom
on the critical behavior of the model, we determine the
universality class of the transition for several values of the number
of colors $N_c$ and of the numbers of flavors $N_f$, also varying the
plaquette-coupling coefficient $\beta_g$. In the two-flavor case, we
always observe a continuous phase transition, in the 3D O(5)
universality class for $N_c=2$ and in the 3D O(3) universality class
for $N_c=3, 4$.  For $N_f=3$ we instead find results compatible with
the presence of a first-order phase transition both for $N_c=2$ and
$3$.

These results agree with the predictions of a LGW analysis based on a
gauge-invariant order parameter \cite{Bonati:2019zrt}, and therefore
indicate the irrelevance of the nonabelian gauge degrees of freedom at
the finite-temperature transition.  In other words, gauge invariance
does not play any role at the transition, apart from that of
restricting the fields to the coset O($M$)/SU($N_c$)
where $M=2N_cN_f$.  Such a
conclusion is also consistent with the observed behavior of the
correlation function $G_V$, defined in Eq.~\eqref{Gv}, which directly
involves the gauge degrees of freedom. In all cases, this correlation
function is short-ranged at the transition.

These results strongly support the procedure initially advocated by
Pisarski and Wilczek in Ref.~\cite{PW-84} to study the chiral phase
transition in massless QCD, which makes use of gauge-invariant order
parameters to analyze the critical behavior of gauge theories when a
global symmetry gets spontaneously broken.

We finally note that there are still several points 
which deserve to be further investigated.
For example, in this work we concentrated on the gauge theory that is obtained 
by starting from a maximally symmetric O($M$)-invariant model and 
by fixing $\mathrm{Tr} Z_{\bm x}^{\dag} Z_{\bm x}=1$. 
It would be interesting to investigate what happens if one or both of these
conditions are relaxed.  It would also be interesting 
to study theories with different global and
local symmetries that are different from the ones considered in this work.

\vspace{0.5cm}

\emph{Acknowledgement} Numerical simulations have been performed on
the CSN4 cluster of the Scientific Computing Center at INFN-PISA. We
thank Daniele Teresi and Omar Zanusso for useful discussions.

\appendix

\section{Symplectic observables for $N_c=2$}
\label{app:symp}

For $N_c=2$ the order parameter is the symplectic analogue of $Q_{\bm
  x}$ defined in Eq.~\eqref{Tdef}.  It is a $2N_f\times 2N_f$
hermitian traceless matrix, which satisfies the relation
\begin{equation}
J\bar{\cal T}J+{\cal T}=0\, , 
\end{equation}
which follows from Eq.~\eqref{condconj}.  It is thus an element of the
$\mathfrak{sp}$($N_f$) algebra \cite{Simon-book}. It can be
parametrized in the block form
\begin{equation}\label{spalgebra}
{\cal T} = \left(\begin{array}{cc} A_1 & A_2 \\ A_3 & A_4\end{array}\right)\ ,
\end{equation}
where $A_1$, $A_2$, $A_3$ and $A_4$ are $N_f\times N_f$ matrices, $A_1$ is 
hermitian and 
traceless, $A_3$ is antisymmetric and 
\begin{equation}
A_4=\bar{A}_1\ ,\qquad A_3=-\bar{A}_2\ .
\end{equation}
It is not difficult to show that ${\cal T}$ can be expressed in terms of
the two order parameters 
$Q_{\bm x}^{fg}$ and $D_{\bm x}^{fg}$. Indeed, we have 
\begin{equation}
A_1=Q\ ,\qquad A_3=-D\ .
\end{equation}
This result implies that the critical behavior encoded in ${\cal
  T}_{\bm x}^{\alpha\beta}$ can be also investigated by studying
$Q_{\bm x}^{fg}$. However, some care should be exercised, when
comparing the results with the Sp($N_f$) predictions.  We define the
correlation function of the ${\cal T}$ field:
\begin{equation}
G_\Gamma({\bm x} - {\bm y}) = 
    \langle \hbox{Tr } {\cal T}_{\bm x} {\cal T}_{\bm y}
   \rangle\; .
\end{equation}
Such a correlation can be related to the correlation of the $Q$ field
defined in Eq.~(\ref{gxyp}). We use the relation
\begin{eqnarray}
&& \langle {\cal T}_{\bm x}^{\alpha\beta} {\cal T}_{\bm y}^{\gamma\delta} 
  \rangle  =
{1\over 2 (N_f-1) (2 N_f+1)} G_\Gamma({\bm x} - {\bm y}) \nonumber \\
&&\quad \times \left(
  J^{\alpha\gamma} J^{\beta\delta} + 
  \delta^{\alpha\delta} \delta^{\beta\gamma} - 
{1\over N_f} \delta^{\alpha\beta} \delta^{\gamma\delta}
    \right),
\end{eqnarray}
which follows from the Sp($N_f$) invariance of the theory. We obtain the
relation 
\begin{equation}
G_\Gamma({\bm x}) = {2 (2 N_f + 1) \over (N_f+1)} G({\bm x})\,.
\end{equation}
It implies that, if one uses Eq.~(\ref{xidefpb}), the same correlation
length is obtained from $G_\Gamma({\bm x})$ or $G({\bm x})$. The
behavior of the Binder parameter is more involved.  For $N_c = 2$ the
natural Binder parameter is
\begin{equation}
U_\Gamma = {\langle \nu_2^2 \rangle \over \langle \nu_2 \rangle^2} \,,
\qquad 
\nu_2 = {1\over V^2} \sum_{\bm x \bm y} \hbox{Tr } {\cal T}_{\bm x}
{\cal T}_{\bm y}\,.
\label{UGammadef}
\end{equation}
In general such a quantity is not related to $U$ defined in
Eq.~(\ref{binderdef}), except for $N_f = 2$, as we discuss below.

For $N_f=2$ the global invariance group is isomorphic to SO(5). It is
useful to make this correspondence explicit. We can rewrite the blocks
$A_1$ and $A_3$ in Eq.~(\ref{spalgebra}) as
\begin{equation}
\begin{aligned}
& A_1=\left(\begin{array}{cc} \phi_3  & \phi_1-i\phi_2 \\ \phi_1+i\phi_2 
& -\phi_3\end{array}\right)\ , \\
& A_3=\left(\begin{array}{cc} 0 & \phi_4 + i\phi_5 \\ -\phi_4-i\phi_5 
& 0\end{array}\right)\ .
\end{aligned}
\end{equation}
Since 
\begin{equation}
({\cal T}^2)^{\alpha\beta} = {1\over 4} \delta^{\alpha\beta},
\end{equation}
we can verify that 
\begin{equation}
\sum_{i=1}^5 \phi^2_i = 1.
\end{equation}
Moreover, one can easily verify that, under infinitesimal Sp(2)
transformations, the vector $(\phi_1,\ldots,\phi_5)$ transforms as an
SO(5) vector. Thus, the redefinition ${\cal T}\to \phi$ explicitly
realizes the isomorphism between Sp(2)$/{\mathbb Z}_2$ and SO(5).
Since
\begin{eqnarray}
&& 
\hbox{Tr} \,Q_{\bm x}Q_{\bm y} = 
  2 \sum_{a=1}^3 \phi^a_{\bm x} \phi^a_{\bm y}\,, \nonumber \\
&& \bar{D}_{\bm x} D_{\bm y} = 
  \sum_{a=4}^5 \phi^a_{\bm x} \phi^a_{\bm y}\,,
\end{eqnarray}
we obtain the relations
\begin{eqnarray}
G_\Gamma({\bm x-\bm y}) &=& 
   4 \langle {\bm \phi}_{\bm x}\cdot {\bm \phi}_{\bm y} \rangle\,, \nonumber  \\
G({\bm x-\bm y}) &=& {3\over 20} G_\Gamma({\bm x-\bm y})\,, \nonumber \\
G_D({\bm x-\bm y}) &=& {1\over 10} G_\Gamma({\bm x-\bm y})\,,
\end{eqnarray}
where we have used the O(5) symmetry of the theory.
For the Binder parameters we have 
\begin{equation}
U_\Gamma = {\langle \nu_{2\phi}^2\rangle \over 
             \langle \nu_{2\phi} \rangle^2}\,, \qquad 
  \nu_{2\phi} = {1\over V^2} \sum_{\bm xy} 
      \langle {\bm \phi}_{\bm x} \cdot {\bm \phi}_{\bm y}  \rangle\,,
\label{UGamma22}
\end{equation}
which shows that $U_\Gamma$ corresponds to the usual O(5) Binder
parameter, and
\begin{equation}
U = {25\over 21} U_\Gamma \,,\qquad 
U_D = {10\over 7} U_\Gamma\,.
\label{app:relationsU}
\end{equation}

\section{Symplectic Landau-Ginzburg-Wilson theory for $N_c=2$}
\label{app:sympLGW}

To define the LGW theory for $N_c=2$, we introduce a  coarse-grained continuum
analogue $\Psi$ of ${\cal T}$, which satisfies the condition 
\begin{equation}
J\bar{\Psi}J+{\Psi}=0\ . 
\end{equation}
The corresponding action is given in Eq.~(\ref{SLGW}). For 
$N_f = 2$, we obtain the O(5) LGW model.
Indeed, in this case we can set
\begin{equation}
\begin{aligned}
& A_1=\left(\begin{array}{cc} \psi_3  & \psi_1-i\psi_2 \\ 
\psi_1+i\psi_2 & -\psi_3\end{array}\right)\ , \\
& A_3=\left(\begin{array}{cc} 0 & \psi_4 + i\psi_5 \\ 
-\psi_4-i\psi_5 & 0\end{array}\right)\ ,
\end{aligned}
\end{equation}
from which it easily follows that 
\begin{equation}
\Psi^2=I\left(\sum_{i=1}^5\psi^2\right)\ ,\quad \mathrm{Tr}\Psi^3=0\,,
\end{equation}
and thus the LGW effective theory for $\Psi$ in the Sp(2) case is
equivalent to that for the O(5) vector model.

\section{The behavior for $\beta\to\infty$}
\label{app:largebeta}

In this appendix we study the large-$\beta$ limit of the model
described by the action $S_g$, Eq.~\eqref{hgauge}.  As the system is
ferromagnetic, the global minimum of the $\beta$-dependent part of the
action is obtained by minimizing the contribution of each link. This
is obtained by setting
\begin{equation}\label{mincond}
Z_{\bm x}=U_{{\bm x},\mu}Z_{{\bm x}+\hat{\mu}}
\end{equation}
on each link. This relation implies that 
\begin{equation}
Q_{\bm x}= Q_{\bm x +\hat{\mu}}\,, \qquad
D_{\bm x}= D_{\bm x +\hat{\mu}}\,, \qquad
\end{equation}
on each link, where $Q_{\bm x}$ and $D_{\bm x}$ are the order
parameters defined in Eqs.~(\ref{qdef}) and (\ref{Ddef}).  The
unit-length condition implies that $Q_{\bm x}$ is nonvanishing: the
system is fully ordered in the limit $\beta \to \infty$ and therefore
the $SU(N_f)$ subgroup is broken at zero temperature. As for the U(1)
order parameter $D_{\bm x}$, we shall show below that $D_{\bm x}$ is
nonvanishing for $N_c = 2$. This is obvious as the U(1) subgroup is a
subgroup of the Sp($N_f$) symmetry group, which is broken. On the
other hand, for $N_c \ge 3$, we find $D_{\bm x}=0$: the U(1) flavor
symmetry is not broken.

Let us now consider any closed path $C_{\bm x}$ that starts and ends
in the same point $\bm x$. By repeated applications of condition
(\ref{mincond}), we obtain the consistency condition
\begin{equation}
Z^{af}_{\bm x} = \sum_{b} 
   \left[ \prod_{l\in C_{\bm x}} U_l\right]^{ab} Z^{bf}_{\bm x}\,.
\label{consistency-cond}
\end{equation}
This relation implies that the product of the links along the path has
at least one unit eigenvalue. For an SU(2) matrix, this implies that
the product is the identity matrix. Therefore, for $N_c =2$, we can
set $U_{{\bm x},\mu} = 1$ modulo gauge transformations. For $N_c \ge
3$, we obtain the condition
\begin{equation}
\prod_{l\in C_{\bm x}} U_l = V^\dagger_{\bm x} W V_{\bm x}
\label{UW}
\end{equation}
with $V_{\bm x} \in \hbox{SU}(N_c)$ and
\begin{equation}
W = \begin{pmatrix}\widehat{W} & 0 \\
                    0 & 1
    \end{pmatrix}\,,
\label{Wblocchi}
\end{equation}
where $\widehat{W}$ is an SU($N_c-1$) matrix. If $\widehat{W}$ does
not have unit eigenvalues (this is the case for a generic unitary
matrix), then
\begin{equation}
  Z_{\bm x} = V^\dagger_{\bm x} A \,,\qquad 
   A = \begin{pmatrix} 0 \\ \hat{z} 
    \end{pmatrix}\,,
\label{Zbetainf}
\end{equation}
where $A$ is an $N_c\times N_f$ matrix such that $A_{ij} = 0$ for any
$i=1,\ldots N_c-1$; $\hat{z}$ is a unit vector of $N_f$ elements.

To obtain more information on the gauge configurations relevant for
$\beta\to\infty$ we have performed simulations for $\beta_g = 0$ on
small lattices ($2^3$ and $4^3$) for very large $\beta$ values (from
$\beta=50$ up to $\beta=300$) and then we have extrapolated the
results to $\beta\to\infty$.  Results for different quantities are
reported in Tables~\ref{l2d3} and \ref{l4d3}. Note that we are indeed
probing the system in the large $\beta$ regime as the average energy
$E$ defined in Eq.~(\ref{ecvdef}) converges to $-3$.  In
Tables~\ref{l2d3} and \ref{l4d3} we also report the average gauge
energy defined in Eq.~(\ref{gauge-energy}).  For $N_c = 2$, results
are consistent with the plaquette being the identity matrix. For $N_c
\ge 3$, data for the average gauge energy, cf. Eq.~(\ref{gauge-energy}),
are consistent with
\begin{equation}
    E_g = {1 \over N_c}.
\label{aveplaquette}
\end{equation}
Note that this is not an exact equality for finite $L$. However,
deviations decrease as $L$ increases from 2 to 4. Such a result can be
explained by assuming that the relevant configurations are such that
all plaquettes can be rewritten in the form (\ref{UW}), where $W$ is
given in Eq.~(\ref{Wblocchi}). Indeed, if this is the case and
$\widehat{W}$ is randomly distributed, we obtain the result
(\ref{aveplaquette}). Of course, we are not claiming that all
minimizing configurations are such that Eq.~(\ref{Wblocchi}) and
(\ref{Zbetainf}) hold.  We only claim that the number of these
configurations is exponentially larger in the lattice volume than the
others, so that they dominate the effective asymptotic behavior. As a
check, we have determined the average of $P^2$, where $P_{{\bm
    x}}^{fg}$ is defined by
\begin{eqnarray}
P_{{\bm x}}^{fg} = \sum_a \bar{Z}_{\bm x}^{af} Z_{\bm x}^{ag}.
\end{eqnarray}
In general, such an operator is not a projector, i.e., $P^2 \not=
P$. However, if the $Z$ fields satisfy Eq.~(\ref{Zbetainf}), we have
$P^2 = P$ and in particular $\hbox{Tr} P^2 = 1$. The results reported
in Tables ~\ref{l2d3} and \ref{l4d3} are in perfect agreement with
this result, confirming the above analysis.

If the relevant configurations have the form (\ref{Zbetainf}) it is
immediate to prove that $D_{\bm x} = 0$ everywhere. The U(1) flavor
symmetry is not broken at $\beta=\infty$, at least for $\beta_g =
0$. It is easy to understand under which conditions the order
parameter $D_{\bm x}$ is not zero.  If we imagine the field $Z^{af}$
as a collection of $N_f$ complex vectors of dimension $N_c$, then
$D_{\bm x}$ does not vanish if $N_c$ of these vectors are nonvanishing
and linearly independent. If this occurs, the consistency condition
(\ref{consistency-cond}) implies that the product of the links along
any path has $N_c$ unit eigenvalues. As the product is an SU($N_c$)
matrix, it must be equal to the identity matrix, which implies that
all gauge fields are equivalent to the identity modulo gauge
transformations. This argument shows therefore that the U(1) symmetry
can be broken only if the relevant configurations are characterized by
the triviality of the gauge fields.  For $N_c \ge 3$ and $\beta_g =
0$, this does not occur and the U(1) symmetry is unbroken. For
$\beta_g = \infty$, there is no gauge dependence and the U(1) symmetry
is broken, an obvious result given that the U(1) group is a subgroup
of the larger O($2N_fN_c$) group. As we expect the gauge energy $E_g$
to depend smoothly on $\beta_g$, we should always have $E_g < 1$ for
finite $\beta_g$: there are relevant nontrivial gauge configurations
that always forbid the breaking of the U(1) symmetry.

Let us finally discuss the behavior for $N_c = 2$. In this case, we
can set $U_{{\bm x},\mu} = 1$ everywhere. Eq.~(\ref{mincond}) implies that
$Z^{af}_{\bm x}$ takes the same value on each link. Thus, the average
of any quantity ${\cal O} (Z_{\bm x})$ can be obtained as
\begin{equation} 
\langle {\cal O}((Z_{\bm x})  \rangle = \int [dA] {\cal O}(A) , 
\end{equation} 
where $A$ is an $N_c\times N_f$ matrix ($N_c = 2$) 
and $[dA]$ is the normalized invariant integration measure over the 
$N_c N_f$-dimensional complex sphere defined by ${\rm Tr} A^\dagger A=1$.

We obtain 
\begin{equation} \label{myguess}
\langle \mathrm{Tr}P_{\bm x}^2\rangle = 
  \int [dA] \;{\rm Tr} [(A^\dagger A)^2] = \frac{N_f + N_c}{1 + N_f N_c}\,,
\end{equation} 
and 
\begin{eqnarray}
U 
 =\frac{(1+N_f N_c) (N_f N_c + 4 N_f^2 + N_f^3 N_c - 6)}{
   (N_f^2-1) (2+N_f N_c) (3+N_f N_c)}\,,
\label{LTvalue}
\end{eqnarray}
which again are consistent with the numerical data in
Tables~\ref{l2d3} and \ref{l4d3} for $N_c=2$. Note that the results
for $U$ are consistent with $U_\Gamma = 1$ when $N_f = 2$, see
Eq.~(\ref{app:relationsU}).

The results that we have obtained for $N_c = 2$ do not depend on the
dimensionality of the system.  On the other hand, for $N_c\ge 3$ the
conclusions we have obtained rely on the fact that the relevant
configurations have the form (\ref{Wblocchi}) and (\ref{Zbetainf}), a
claim that is only justified by the numerical study we have performed
on cubic lattices. We expect, but we do not have a proof, that the
same result holds in any dimension.

\begin{table*}
\begin{tabular}{cllllll}
\hline\hline
$(N_c, N_f)$ & \multicolumn{1}{c}{$E_g$}   & 
    \multicolumn{1}{c}{$E/3$}  &   \multicolumn{1}{c}{$U$}   &
Eq.~\eqref{LTvalue}   & $\mathrm{Tr }\, P^2$ & Eq.~\eqref{myguess}   \\ \hline
(2, 2)       & 0.99998(2)  & $-$0.999992(6)  & 1.191(2)     & 1.19048$\ldots$ & 0.800(1) & 0.8                  \\ 
(2, 3)       & 1.00000(1)  & $-$1.000010(5)  & 1.0940(5)    & 1.09375$\ldots$ & 0.714(2) & 0.714286$\ldots$     \\
(2, 4)       & 1.00001(1)  & $-$1.000000(4)  & 1.0582(3)    & 1.05818$\ldots$ & 0.666(1) & 0.666667$\ldots$     \\
(2, 5)       & 1.00000(1)  & $-$1.000000(6)  & 1.0400(3)    & 1.04006$\ldots$ & 0.636(1) & 0.636364$\ldots$     \\
(2, 6)       & 0.999990(6) & $-$1.000000(4)  & 1.0294(2)    & 1.02939$\ldots$ & 0.615(2) & 0.615385$\ldots$     \\ \hline

(3, 2)       & 0.3347(3)   & $-$1.000005(6)  & 1.0000000(1) &                 & 1.00001(1) &                  \\
(3, 3)       & 0.3369(4)   & $-$0.999993(6)  & 1.0000000(6) &                 & 0.99999(1) &                  \\
(3, 4)       & 0.3382(4)   & $-$1.000000(7)  & 1.0000000(6) &                 & 1.00000(1) &                  \\
(3, 5)       & 0.3413(4)   & $-$1.000000(5)  & 1.0000000(1) &                 & 1.00000(1) &                  \\
(3, 6)       & 0.3463(4)   & $-$1.000000(5)  & 1.00000000(4)&                 & 0.99999(1) &                  \\ \hline

(4, 2)       & 0.2500(2)   & $-$0.99999(1)   & 1.0000000(4) &                 & 1.00000(1) &                  \\
(4, 3)       & 0.2510(2)   & $-$1.00001(1)   & 1.0000000(2) &                 & 1.00001(2) &                  \\
(4, 4)       & 0.2513(3)   & $-$0.99995(4)   & 1.0000000(3) &                 & 0.99992(8) &                  \\
(4, 5)       & 0.2520(4)   & $-$1.000010(7)  & 1.0000000(1) &                 & 1.00002(2) &                  \\
(4, 6)       & 0.2527(3)   & $-$1.000000(6)  & 1.0000000(1) &                 & 0.99999(1) &                  \\ 
\hline\hline
\end{tabular}
\caption{Asymptotic values for $\beta\to\infty$ on a $2^3$ lattice at $\beta_g=0$.}
\label{l2d3}
\end{table*}

\begin{table*}
\begin{tabular}{cllllll}
\hline\hline
$(N_c, N_f)$ & \multicolumn{1}{c}{$E_g$}  & \multicolumn{1}{c}{$E/3$}   & 
    \multicolumn{1}{c}{$U$}  &   
   Eq.~\eqref{LTvalue}  & $\mathrm{Tr }\, P^2$ &  Eq.~\eqref{myguess}  \\ \hline
(2, 2)       & 0.99998(2)  & $-$0.999999(2)  & 1.195(3)      & 1.19048$\ldots$ & 0.798(1)  & 0.8               \\ 
(2, 3)       & 1.000000(6) & $-$1.000000(2)  & 1.094(1)      & 1.09375$\ldots$ & 0.714(3)  & 0.714286$\ldots$  \\
(2, 4)       & 0.999999(3) & $-$1.000000(3)  & 1.0580(5)     & 1.05818$\ldots$ & 0.666(2)  & 0.666667$\ldots$  \\
(2, 5)       & 1.000000(3) & $-$1.000000(3)  & 1.040(1)      & 1.04006$\ldots$ & 0.637(2)  & 0.636364$\ldots$  \\
(2, 6)       & 1.000000(3) & $-$0.999988(5)  & 1.0293(5)     & 1.02939$\ldots$ & 0.615(2)  & 0.615385$\ldots$  \\ \hline
                                                                                                                                           
(3, 2)       & 0.3344(2)   & $-$1.000000(3)  & 1.000000(3)   &                 & 1.000000(6) &                   \\
(3, 3)       & 0.3356(2)   & $-$1.000000(3)  & 1.000000(3)   &                 & 0.999999(6) &                   \\
(3, 4)       & 0.3368(2)   & $-$0.999999(3)  & 1.000000(3)   &                 & 1.000008(8) &                   \\
(3, 5)       & 0.3375(2)   & $-$1.000000(3)  & 1.00000000(5) &                 & 1.00001(1)  &                   \\
(3, 6)       & 0.3385(2)   & $-$0.999996(4)  & 1.00000000(2) &                 & 0.99999(1)  &                   \\ \hline
(4, 2)       & 0.2501(2)   & $-$1.00000(1)   & 1.0000000(1)  &                 &
0.99999(2) &                   \\
\hline\hline
\end{tabular}
\caption{Asymptotic values for $\beta\to\infty$ on a $4^3$ lattice at $\beta_g=0$.}
\label{l4d3}
\end{table*}

\end{document}